\documentclass[10pt]{article}
\usepackage[utf8]{inputenc}
\usepackage[centertags]{amsmath} 
\usepackage{amssymb} 
\usepackage[psamsfonts,mathscr]{eucal} 
\usepackage{ifthen,calc,enumerate}
\usepackage[all,frame,import]{xy}
\usepackage{graphicx}
\usepackage{subfigure}
\usepackage[comma,numbers,sort&compress]{natbib}
\usepackage{lineno}
\usepackage{wrapfig,lipsum,booktabs}
\usepackage{enumerate}
\usepackage{authblk}
\usepackage{caption}
\usepackage{pifont}
\usepackage{booktabs}
\usepackage{nomencl}
\usepackage[toc,page]{appendix}
\usepackage[table,xcdraw]{xcolor}
\usepackage[T1]{fontenc}
\makenomenclature
\usepackage[%
     left=0.85in,%
     right=0.85in,%
    top=0.75in,%
    bottom=1.0in,%
    paperheight=11in,%
    paperwidth=8.5in%
]{geometry}%

\usepackage{tikz}
\usepackage{xcolor}
\usetikzlibrary{arrows,arrows.meta,patterns,intersections,calc,decorations.pathmorphing,decorations.markings} 
\usetikzlibrary{positioning}
\tikzstyle{dim}    = [latex-latex]

\newcommand{\abs}[1]{\left\vert#1\right\vert}

\newcommand{\eq}[1]{Eq.\ (\ref{#1})}

\newcommand{\fig}[1]{Fig.~\ref{#1}}

\newcommand{\figtwo}[2]{Figs.~\ref{#1} and~\ref{#2}}

\newcommand\blfootnote[1]{%
  \begingroup
  \renewcommand\thefootnote{}\footnote{#1}%
  \addtocounter{footnote}{-1}%
  \endgroup
}

\graphicspath{{./figures/}} 
\providecommand{\keywords}[1]{\textbf{Keywords---} #1}

\begin{document}
\font\titlefont=cmr12 at 14 pt
\font\authorfont=cmr12 at 12 pt


\title{\bf Fatigue Life Estimation of Structures under Statistically and Spectrally Similar Variable 
Amplitude Loading}
\author{\authorfont He-Wen-Xuan Li and David Chelidze\footnote{Corresponding author.\\
E-mail address: chelidze@uri.edu (D.Chelidze).\\
URL:http://web.uri.edu/engineering/meet/chelidze (D.Chelidze).}}
\affil{\authorfont Nonlinear Dynamics Laboratory\\
Department of Mechanical, Industrial and Systems Engineering\\
University of Rhode Island, Kingston, RI 02881\\
}
\date{}

\maketitle


\begin{abstract}
A new fatigue life prediction framework provides an improved life prediction under statistically and spectrally similar irregular variable-amplitude loading for a notched beam model.
It enables the cumulative damage rule to account for the load sequence effects by modifying the probability density function of the stress-amplitude history through (1) identification of overloads based on the rainflow-counting algorithm; (2) analytical characterization of the overload retardation effects; and (3) correction to the damage rule using overload amplitude rate characterization. 
The fatigue lives estimated from experimentally acquired and synthetically generated load-time histories are compared to the ones generated from simulations that qualitatively reproduce the fatigue lives in physical experiments. 
The notable improvement in prediction accuracy outperforms the Palmgren-Miner's rule and power-spectrum-based life estimation. 
The demonstrated application to the field acceleration data substantiates its use for in-service structural health monitoring and damage prognosis. This framework does not require \emph{a priory} knowledge of the applied load, and it can be applied to other engineered structures with known structural and defect properties. 
\end{abstract}

\blfootnote{\emph{Manuscript submitted to Mechanical Systems and Signal Processing}}

\section{\textbf{Introduction}}
Engineering fatigue is a very prevalent and dangerous phenomenon that limits the useful life span of mechanical structures~\cite{suresh1998fatigue, schijve2001fatigue}.
It can be found in most engineered structures and machinery ~\cite{seidel2014wave,schijve2001fatigue,campbell1984survey,kang2011corrosion,reifsnider2012fatigue,mickens2003structural,du2020damage, zhu2020hybrid, chen2018underwater} that are made from metallic or composite materials.
The life prediction methodology under {\em variable amplitude loading} (VAL), however, has not reached maturity due to the complexity of the loading~\cite{schijve2001fatigue,xu2017fatigue}.
This complexity results in load sequence effects, which significantly alter the fatigue life~\cite{schijve2001fatigue,skorupa1998load}.
Since most of the realistic loading experienced by engineered structures is of variable amplitude nature, and the corresponding responses at the defect site are highly irregular, characterization of the VAL and its resulting damage is paramount.
Fatigue life estimation under VAL conventionally uses methods based on the Miner's Rule or the  {\em cumulative damage rule} (CDR)~\cite{miner1945cumulative,santecchia2016review} in conjunction with the rainflow counting method~\cite{dirlik1985application,amzallag1994standardization}.
However, CDR does not account for the load interaction/sequence effects induced by local residual stress (or crack opening stress) variations~\cite{skorupa1998load,skorupa1999load}.
Although CDR is extensively used in the industry/academia during the design phase for its simplicity, its inability to consider the load sequence effects results in inaccurate damage estimation even for simple VAL~\cite{schijve2001fatigue}.
Modifications based on the CDR are proposed during the past decades~\cite{marco1954concept,manson1986re}.
However, they have limited practical use when the applied load is highly irregular since the stress amplitudes' characterization becomes cumbersome from experimental data.
If statistics of the loading is known, CDR can be described based on the stress amplitude statistics, 
\begin{equation}\label{eq:ettf}
    T = \frac{1}{\nu^+C^{-1}\int_0^{\infty} \sigma_a^k p(\sigma_a)d\sigma_a},
\end{equation}
where $\nu^+$ is the expected rate of occurrence of mean up-crossing for a given loading or power spectrum; $C^{-1}\sigma_\mathrm{a}^k$ is from the corresponding S-N curve (with $k$ and $C$ the slope of the S-N curve and the fatigue strength, respectively); $p(\sigma_\mathrm{a})$ is the \emph{probability density function} (PDF) of the applied stress amplitude history $\sigma_\mathrm{a}$ obtained from cycle counting.
Connections between the PDFs and the \emph{power spectral density} (PSD) functions are established and are referred to as the spectral CDR~\cite{rice1944mathematical,miles1954structural,john1954structural,wirsching1980fatigue,dirlik1985application,tovo2002cycle,benasciutti2005spectral}.
Applications for the spectral methods in the design phase of mechanical structures can be found in~\cite{zhao1992probability,mrvsnik2013frequency}.
However, to the authors' best knowledge, estimating fatigue life from a mechanical system whose response/loading is non-Gaussian is not within reach of the spectral methods.
If the structure's damage is detected, elastoplastic-fracture-mechanics-based \emph{fatigue crack propagation} (FCP) method can be used for damage prognosis.
The number of cycles needed to reach critical-fatigue-crack size describes the fatigue life. 
Based on the Paris' Law, these methods correlate the crack propagation rate to the applied-stress-intensity-factor range.
The FCP methods model the crack propagation rate as a functional form of the applied effective-stress-intensity-factor range~\cite{schijve2001fatigue,sanford2003principles,Harter2019AFGROWUG}, 
\begin{equation}
\frac{da}{dN} = f(\Delta K_\mathrm{eff})    
\end{equation}
where $\Delta K_\mathrm{eff} = g(\sigma, R, F_g, \boldsymbol{\mu})$ is a function of the stress $\sigma$, the $R$-ratio (defined as ${R} = \sigma_\mathrm{min}/\sigma_\mathrm{max}$), the geometry of the defect $F_g$, and a set of material parameters $\boldsymbol{\mu}$.
The total number of \emph{cycles-to-failure} (CTF) can be obtained by integrating the differential equation iteratively.
This formulation ensures the description of load sequence effects by incorporating a detailed description of the loading, compared to the CDR.
Shortcomings of this method include the difficulty in obtaining the proper material parameters that characterize load interaction effects and the need for more extensive computational resources than the CDR (which needs a statistical description of the loading signal).
Moreover, when the actual load time history or the appropriate tracking of the defect size is not available, the FCP method cannot be applied.
Therefore, for \emph{structural health monitoring} (SHM), the CDR methods become unreliable whenever the applied loads deviate from Gaussianity or when the signals are the combination of determinism and randomness.
The FCP methods, although comparatively accurate compared to the CDR, require explicit tracking of the load-time history and the damage variables.

Accurate fatigue analysis and the evaluation of structural degradation depend on the proper setup of SHM system and suitable data processing techniques, which remain open research areas~\cite{yuan2017weak,stawiarski2017fatigue, shamsudin2019application,zhang2018approach, chen2020data}.
SHM mainly focuses on load identification, damage diagnostics, and damage prognosis in structural and mechanical systems by analyzing field measurements~\cite{adams2007health,farrar2012structural}.
Therefore, characterization of the applied load and modeling of the damage from the acquired data is the basis of SHM, aside from the advancing sensing technologies.
The load experienced by most designed structures is usually deterministic, and the corresponding response follows the designed behavior.
As the system degrades, its response becomes irregular compared to their healthy state.
Randomness is usually less dominant and is induced by uncertainties in the system parameters (e.g., the imperfection of the machinery or randomness in the external loads); or observed from the collected data as additive noise.
However, it is impossible to differentiate between the deterministic chaotic behavior and the wide-band stochastic one in certain situations.
As a result, CDR cannot yield accurate fatigue life estimates if the statistical and probabilistic descriptions are nearly identical.
Nguyen and Chelidze~\cite{nguyen2013fatigue, nguyen2017dynamic} conducted experiments to explore the fatigue lives under statistically and spectrally similar deterministic and stochastic loads.
The resulting averaged {\em time to failures} (TTF) under chaotic excitation was twice as long as the one under the corresponding random surrogate. 
Moreover, the CDR overestimated the damage to the structure done by the chaotic loading.
Linear statistics (i.e., mean estimate and second-order central moment of the probability mass function of stress level histories) showed no difference between the deterministic (i.e., chaotic) and the stochastic (i.e., surrogate)  load time histories. 
Therefore, methods using only the linear statistics from the univariate distribution of stress history and power spectrum led to erroneous predictions.
Instead, these differences were characterized by the local divergence rate (similar to the largest Lyapunov exponent, an invariant measure of nonlinear dynamical systems~\cite{kantz2004nonlinear}) that reflects the divergence of two initially closely spaced trajectories in the phase space~\cite{nguyen2013fatigue}.
However, the estimation of such nonlinear factors is conceptually challenging and computationally expensive for real-time monitoring~\cite{chen2015condensation}.
Besides, the quantification of fatigue life applied to the considered loading was not based on absolute measurements.
Nevertheless, it provided insights into the underlying governing factors contributing to the difference in the resulting fatigue life.
Therefore, an advanced tool is required to differentiate and characterize damage under such loading.

Aside from structural degradation, new requirements for lightweight, high efficiency, multi-scale, and multi-functional engineered structures require consideration, exploration, and utilization of material and structural nonlinearities. 
The response of a generic nonlinear system can be deterministic but complex in its temporal structure. 
Also, its frequency content is often not limited to distinct components or narrow frequency bandwidth.
For example, the applications of bistable \emph{microelectromechanical systems} (MEMS) to sensors and actuators work under oscillation with multiple stable equilibria, causing multi-modal distribution in stress amplitudes~\cite{saif2000tunable}.  
Coexisting responses and stochastic resonances are commonplace in both discrete systems (e.g., the Duffing's oscillator) and elastic continua (e.g., post-buckled beam) when external forcing is a combination of harmonic and stochastic signals~\cite{wiebe2014co,kim2020numerical}.
The rapidly advancing research field of metamaterials for engineering structures provides more intriguing examples of irregular behaviors.
They can achieve multistable behaviors, i.e., having distinct mechanical responses in each stable state~\cite{kadic20193d,frenzel2016tailored,coulais2016combinatorial}.
Thus, these structures can be multi-functional and, at the same time, prone to the shifts between designed modes.
These irregular loading/responses require modifications to the damage evaluation techniques to reflect the underlying temporal dynamics of loads to account for their non-normality and nonlinearity to obtain reliable fatigue life predictions.

\begin{figure}
    \centering
    \includegraphics[width = 0.7\textwidth]{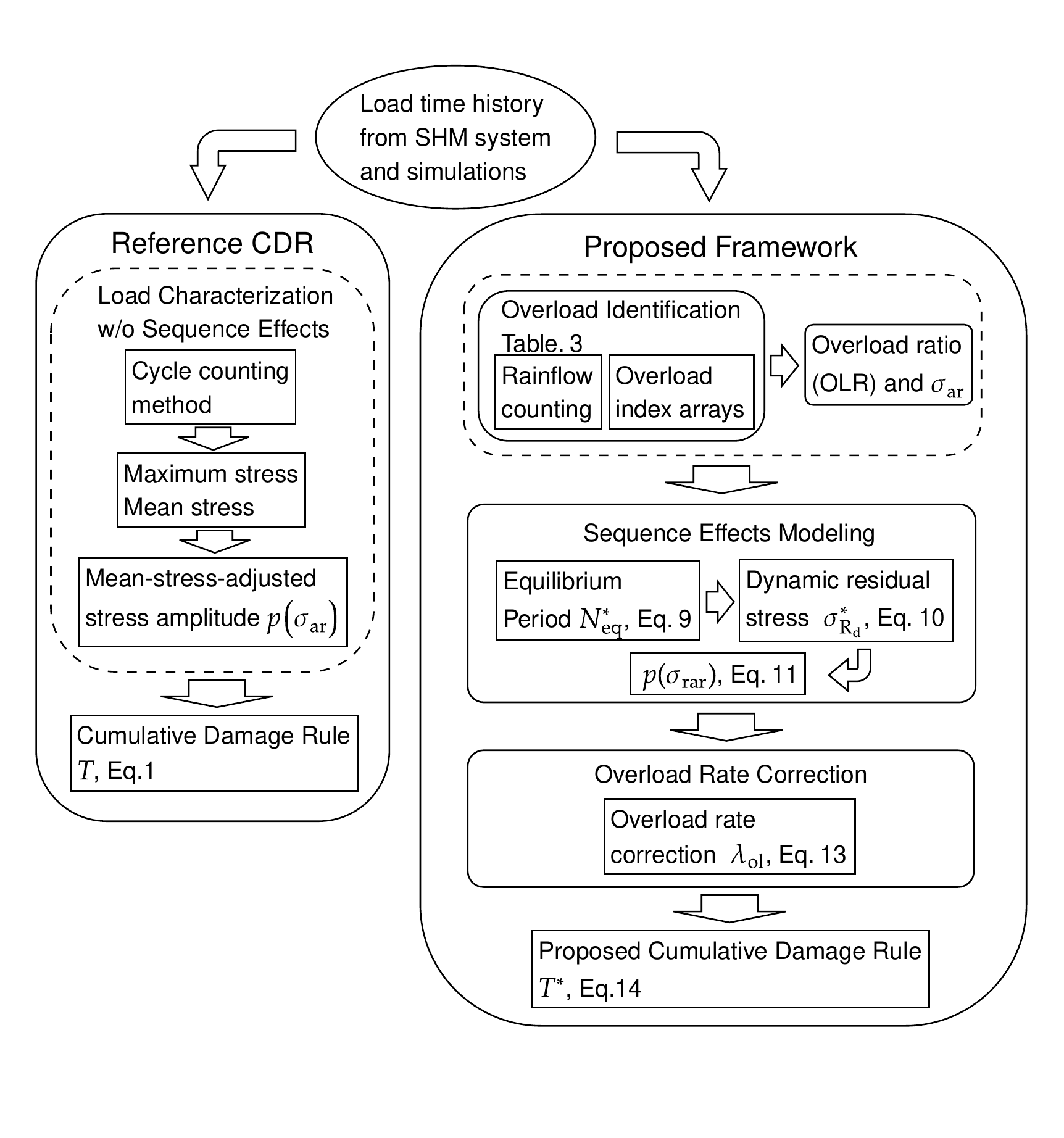}
    \caption{Comparison of workflows between conventional CDR and the proposed framework}
    \label{fig:comp_workflow}
\end{figure}

To overcome the challenges mentioned above, we propose a new and improved framework---built on the observations from physical experiments and numerical simulation---for the accurate fatigue damage estimation. 
The framework modifies the CDR by considering the load sequence effect as weights to the \emph{probability density function} (PDF) of mean-stress-corrected stress amplitudes.
This framework helps delineate the observed difference in fatigue damage dynamics and predicts the expected fatigue life under irregular VAL with similar stress PDFs and PSDs from the field measurement.
The framework incorporates the CDR with the FCP by introducing a novel overload identification and characterization procedure.
First, this framework identifies the location of overloads for a given load time history based on the rainflow-counting algorithm.
Second, a local stress variation function models the effect of crack retardation.
This function is applied to the mean-stress-adjusted stress amplitude based on the cycle range counting.
It models the local residual stress (or equivalently, the crack opening stress) variation, which acts as a temporal weight to the original load time-history.
The local stress variation function adjusts stress amplitudes affected by overloads to the \emph{retarded stress amplitude} (RSA).
The characterization of RSA does not require,  cycle-by-cycle, explicit tracking of the residual stress.
To further consider the effects of underlying overload dynamics, a data-driven correction factor is also incorporated to help characterize the overload amplitude PDF during the damage estimation. 
Finally, the integration of a modified time-to-failure equation provides the estimated fatigue life using the PDF of the RSA.
A comparison of the traditional CDR and the proposed framework is illustrated in a flow chart in \fig{fig:comp_workflow}.

The rest of the paper discusses the proposed framework's details and its application to the two groups of cycles-to-failure data. Section 2 provides the details of the generation of load time histories used in this study and the justification for the simulations that qualitatively capture the fatigue crack propagation dynamics from a prior experimental work~\cite{nguyen2013fatigue,nguyen2017dynamic}.
Section 3 focuses on the formulation of the proposed framework for fatigue life prediction under generic irregular loading. 
Finally, section 4 discusses the proposed framework's capability based on the improved fatigue life prediction results through a comparison between the new framework and other existing methods~\cite{tovo2002cycle,benasciutti2005spectral}. Concluding remarks and future work are given in Section 5.

\section{Load Time Histories Generation and Validation of the Simulation}

Before introducing the framework, a brief introduction to the data used in this study is needed.
As a proxy to the previous experimental work, we have established a simulation environment to facilitate fatigue life investigations under various statistically and spectrally similar irregular loadings.
This numerical simulation environment uses an existing crack closure retardation model, an FCP method with only one predetermined material parameter.
Despite its simplicity, it captured the dynamics of crack propagation and is comparable to the experimental observation in the previous studies~\cite{nguyen2013fatigue,nguyen2017dynamic}.
Apart from time savings, this environment allows investigation the fatigue life under more sophisticated loading, which is not easily realizable in our experimental setup~\cite{falco2014nonlinear}. 
We consider two groups of loading.
The first group is the experimental cycles-to-failure from the previous study, where the input to the mechanical system is the velocity solution of Duffing's oscillator and its random surrogate~\cite{nguyen2013fatigue,nguyen2017dynamic}.
The second group is the simulated cycles-to-failure based on the crack closure model using four chaotic cases and their surrogates.
For brevity, the name ``synthetic loading'' is used for this loading group.
The proposed framework for fatigue life prediction uses the observation from these two groups of fatigue crack propagation dynamics.

This paper categorizes the wide-band VA load time histories as either deterministic or stochastic. 
Each deterministic time history has the corresponding stochastic surrogate loading with similar spectral and statistical characteristics to investigate the differentiating factors resulting in distinctive fatigue lives.
The spectrally and statistically similar chaotic and surrogate loads---the chaotic-surrogate-pairs---are representatives of two types of vibration problems, one being the response of linear mechanical systems under irregular loading~\cite{meirovitch2010fundamentals, lin1967probabilistic}, the other is the response of nonlinear mechanical systems under simple periodic loading or irregular excitation~\cite{Nayfeh1995,nayfeh2008nonlinear,foong2003chaos}.
We also assume direct application of the generated load time histories as the far-field stress near the defect site.
The deterministic loading is typical following the designed operating conditions for the machinery and structures under investigation. 
Periodic excitation is standard for most machinery (e.g., bearings in wind turbines or electric motors and connecting rods in combustion engines.). Aperiodic or even chaotic response occurs when the system itself is nonlinear and has multistable equilibria or is damaged~\cite{kim2020numerical,wei2019analysis,shankar2016nonlinear,foong2003chaos}.
Finally, the stochastic loading happens whenever the system undergoes random vibration (e.g., seismic or wind loading) or when the system has uncertainties in its parameters or boundary conditions~\cite{hu2018laplace,cao2020global}.
In this study, discrete chaotic solutions to well-known chaotic dynamical systems generate the examined chaotic loads.
We randomly select 20 unique equal-length segments of the original long chaotic time series to generate the corresponding stochastic counterparts for each chaotic case using surrogate data analysis. 
The chaotic and surrogate pairs have similar PSD and stress level PDF.

%
\subsection{Generation of the Chaotic Loading}
 
One group of the chaotic responses are generated from sampled numerical solutions of the double-well Duffing's equation~\cite{hilborn2000chaos, sprott2003chaos}
\begin{equation}
    \ddot{x} + 0.25\, \dot{x} - 0.6\, x + x^3 = 0.2\, \mathrm{cos}(t),
\end{equation}
where $t$ is the time variable, $\dot{x}$ is the first order time derivative of the variable $x$.
The double-well Duffing's equation is a typical nonlinear system with two stable and one unstable static equilibrium points.
In addition, the Lorenz equation~\cite{hilborn2000chaos, sprott2003chaos}
\begin{equation}
\begin{cases}
    \dot{x} = 10\,(y - x),\\
    \dot{y} = -xy + 28\,x - y,\\
    \dot{z} = xy - \frac{8}{3}z.
    \end{cases}
\end{equation}
is added to expand the scope of the proposed method's applicability. 
Although the Lorenz equation models the atmospheric convection~\cite{lorenz1963deterministic}, its distinct dynamical characteristics is of our interest.
The first state variable from the Lorenz equation is selected due to its double-well nature; therefore, it can be compared to the Duffing's double-well solution.
Since the third state variable resembles single-well oscillations, and the temporal dynamics is less irregular compared to other cases, it is also included.
We named the chaotic load time histories according to the corresponding nonlinear dynamical system and the corresponding generalized coordinate to simplify the expression.
Thus, the synthetic load-time history obtained from the first coordinate (i.e., the displacement, $x$) and the second coordinate (i.e., the velocity, $\dot{x}$) of the Duffing's oscillator are labeled D1 and D2, respectively.
The other two cases are named as L1 and L3, respectively.

\begin{table}[]
\centering
\begin{tabular}{@{}llllll@{}}
\toprule
                Loading & Mean (MPa) & $\mathrm{Std}$ (MPa) & RMS (MPa) & Skewness        & Kurtosis    \\ \midrule
\rowcolor[HTML]{EFEFEF} 
D1\qquad     & 25                & 70               & 26.36 & 0.087 & 1.82 \\
D1S\qquad & 25                & 70               & 26.36 & 0.087 & 1.82 \\
\rowcolor[HTML]{EFEFEF} 
D2\qquad     & 25                & 70               & 26.36 & -1.7e-3  & 1.90 \\
D2S\qquad & 25                & 70               & 26.36 & -1.6e-3  & 1.91 \\ \midrule
\rowcolor[HTML]{EFEFEF} 
L1\qquad      & 25                & 70               & 26.36 & 0.062     & 2.29 \\
L1S\qquad  & 25                & 70               & 26.36 & 0.062     & 2.29 \\
\rowcolor[HTML]{EFEFEF} 
L3\qquad      & 25                & 70               & 26.36 & 0.191     & 2.17 \\
L3S\qquad  & 25                & 70               & 26.36 & 0.191     & 2.17 \\ \bottomrule
\end{tabular}
\caption{Statistics of the generated synthetic load-time histories used in the simulations}
\label{tab:statistics}
\end{table}

\subsection{Generation of the Surrogate Stochastic Loading}

After obtaining each chaotic load time history, its stochastic surrogate is generated~\cite{schreiber2000surrogate, kantz2004nonlinear}.
Generally, surrogate data analysis tests the nonlinearity of a given set of time series using the null hypothesis.
Here, surrogate data generation serves the purpose of keeping identical spectral magnitude and temporal amplitude statistics between chaotic time histories and their random surrogates.
One can consider this process as varying the phase information of the deterministic loading while keeping the temporal and spectral magnitude information unchanged.
The procedure for the generation of surrogate stochastic loading is as follows.
Step one: randomly permute the amplitudes of the sampled synthetic chaotic load time histories (stress level).
This randomization preserves the stress-level PDF but alters the frequency content (reflected in the power spectrum).
Step two:  apply fast Fourier transform (FFT) to the randomized load-time histories and the original chaotic time series.
Perform a power match-up between the two frequency spectra to ensure a nearly identical frequency content before and after the randomization.
Time series with identical frequency magnitude, later on, are obtained through inverse fast Fourier transform (IFFT).
This step alters the stress-level probability-mass-function slightly.
Iterate these two steps until a good match is obtained in both probabilistic and spectral sense.
The algorithm is given in appendix A.1.
The generated surrogate load-time histories are named according to their chaotic counterparts, with an additional `s' as a suffix in each case (e.g., D1s).

\begin{figure}[t]
    \centering
        \includegraphics[width = 0.45\textwidth]{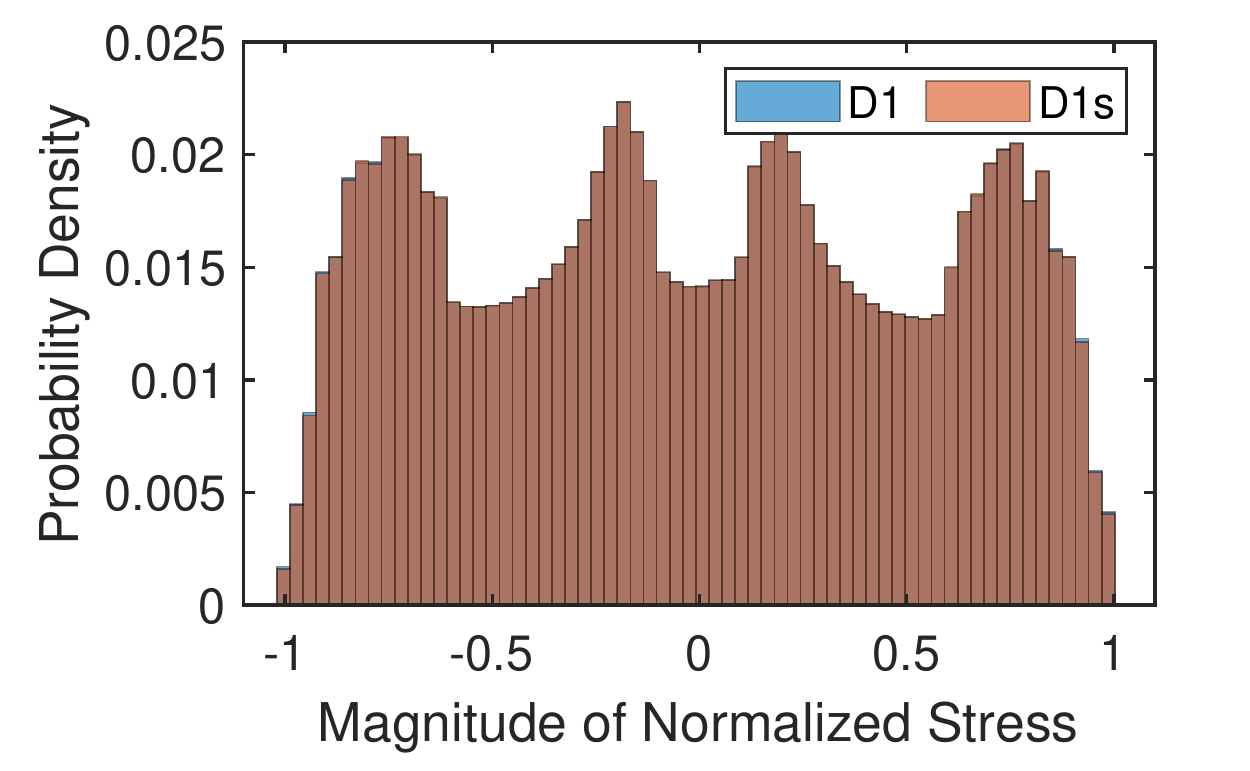}
        \includegraphics[width = 0.45\textwidth]{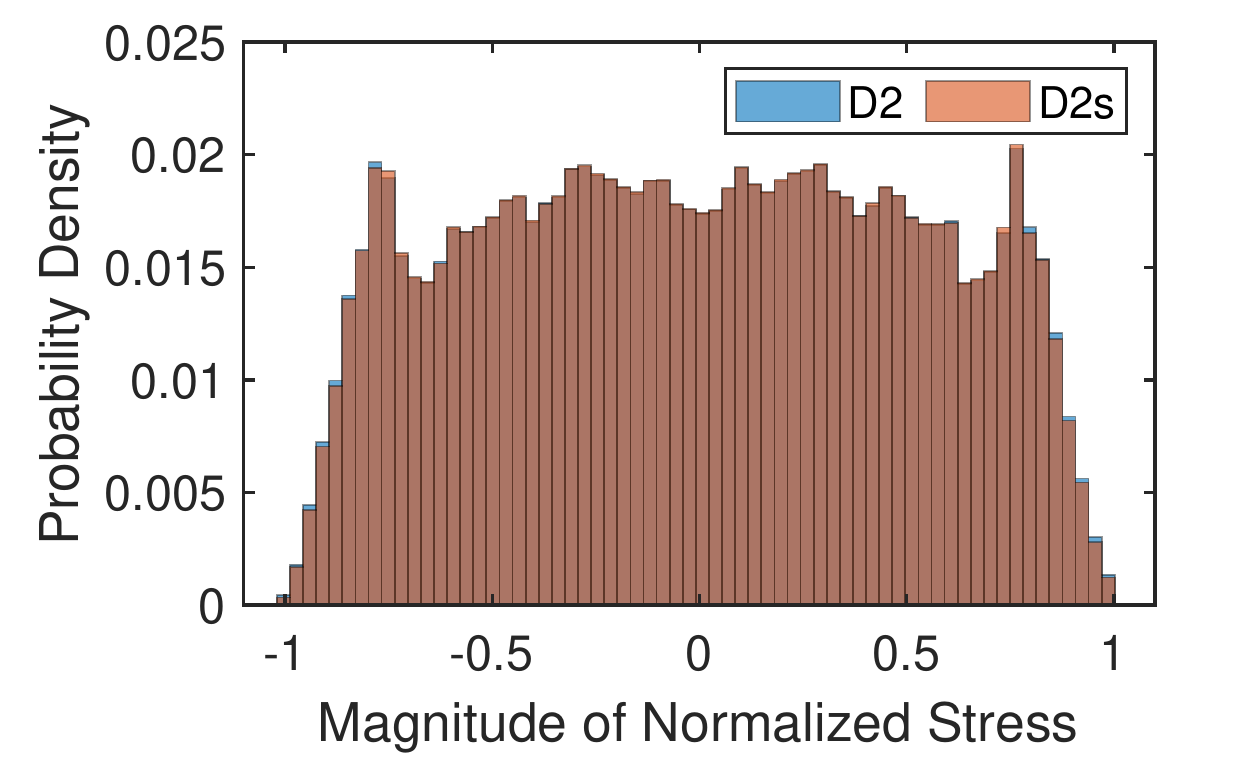}
    \caption{Estimated stress amplitude PDFs indicate similar load distribution for the generated chaotic loading and their surrogates. The PDFs of the chaotic--surrogate loading pairs are on top of each other, indicating nearly identical statistical representation of the two distinct loading. Left: first solution from Duffing's oscillator and its surrogate; right: second solution from Duffing's oscillator and its surrogate.}
    \label{fig:PDF_loading}
\end{figure}

\begin{figure}
    \centering
        \includegraphics[width = 0.275\textwidth]{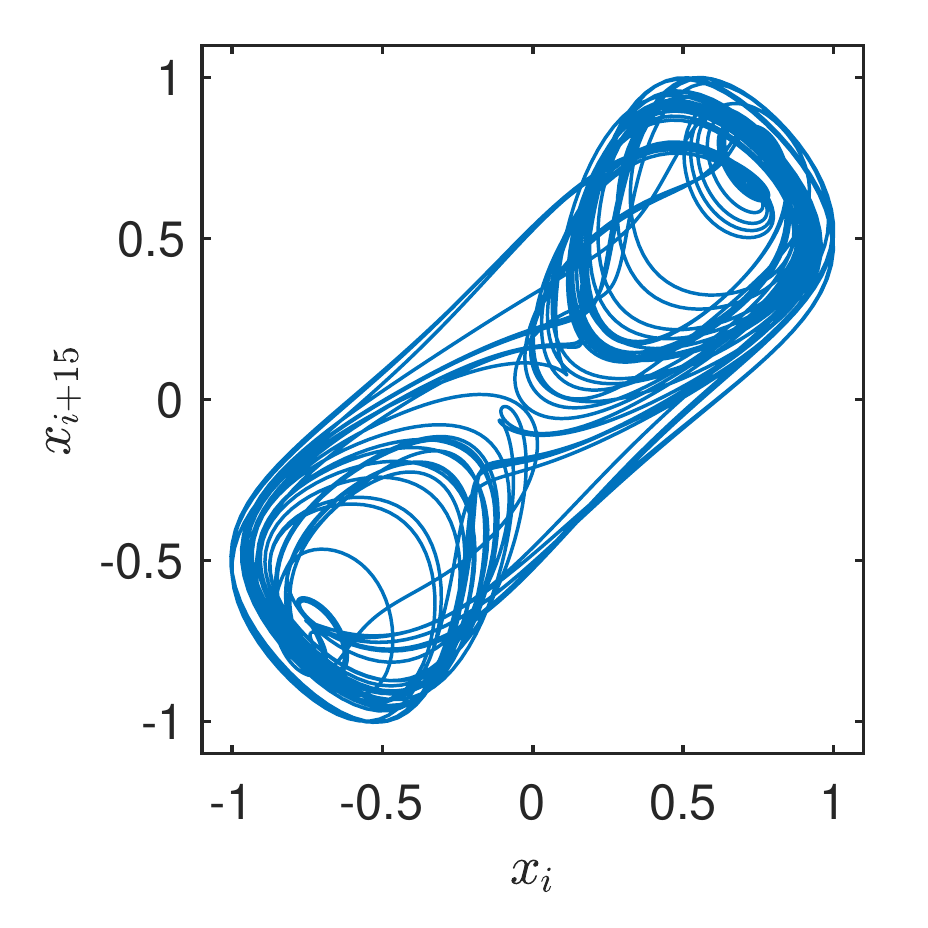}
                \includegraphics[width = 0.45\textwidth]{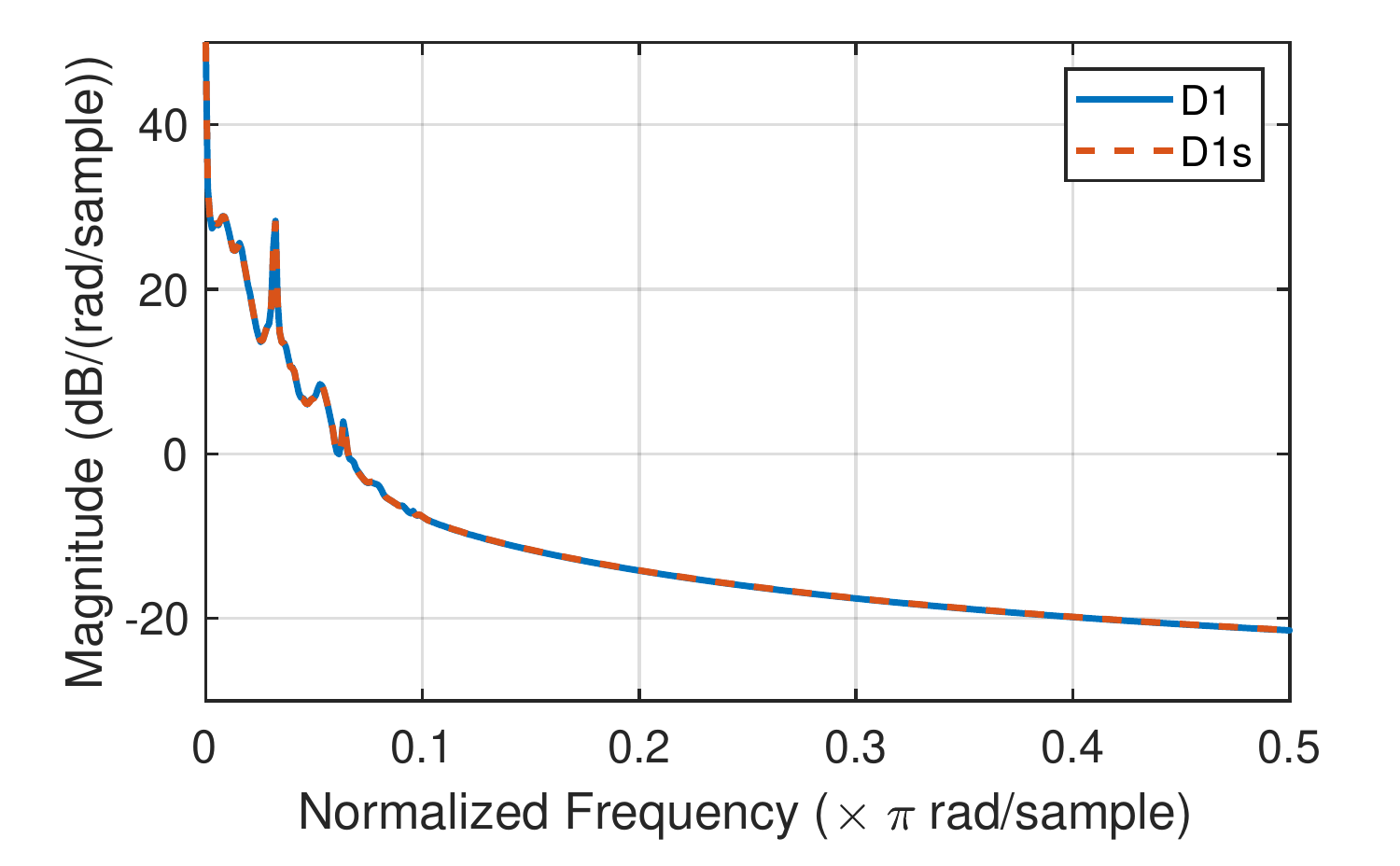}\\
        \includegraphics[width = 0.275\textwidth]{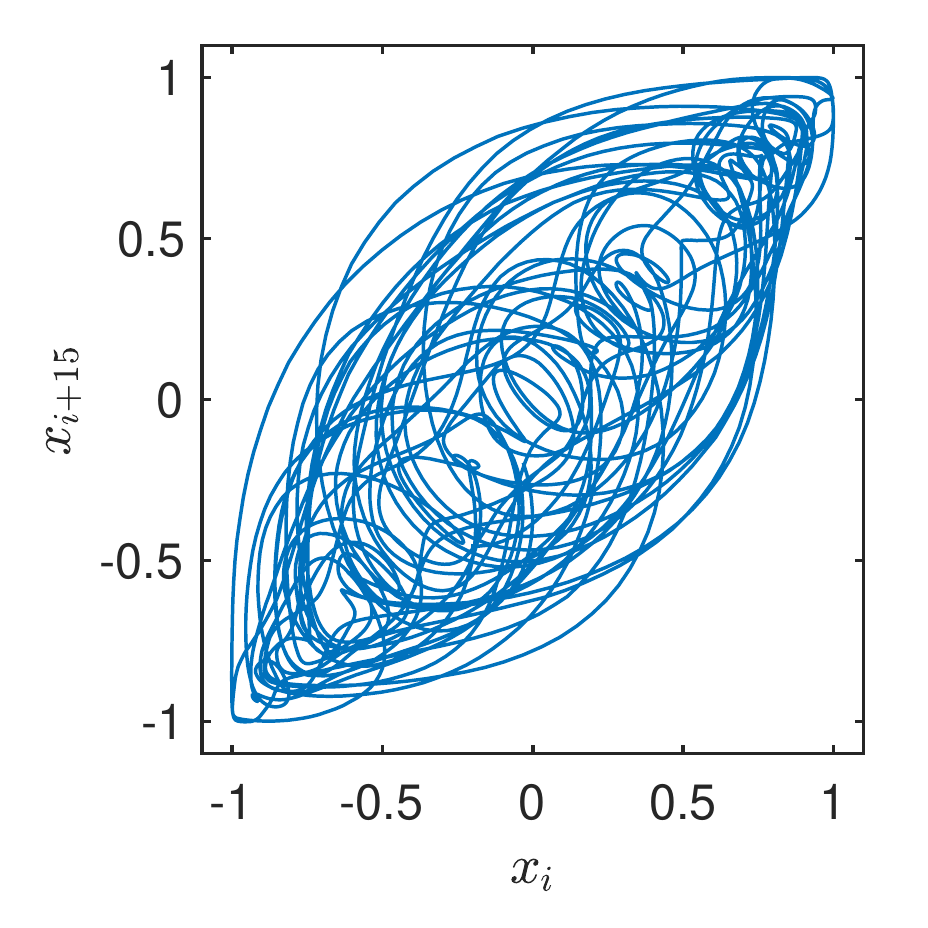}
                \includegraphics[width = 0.45\textwidth]{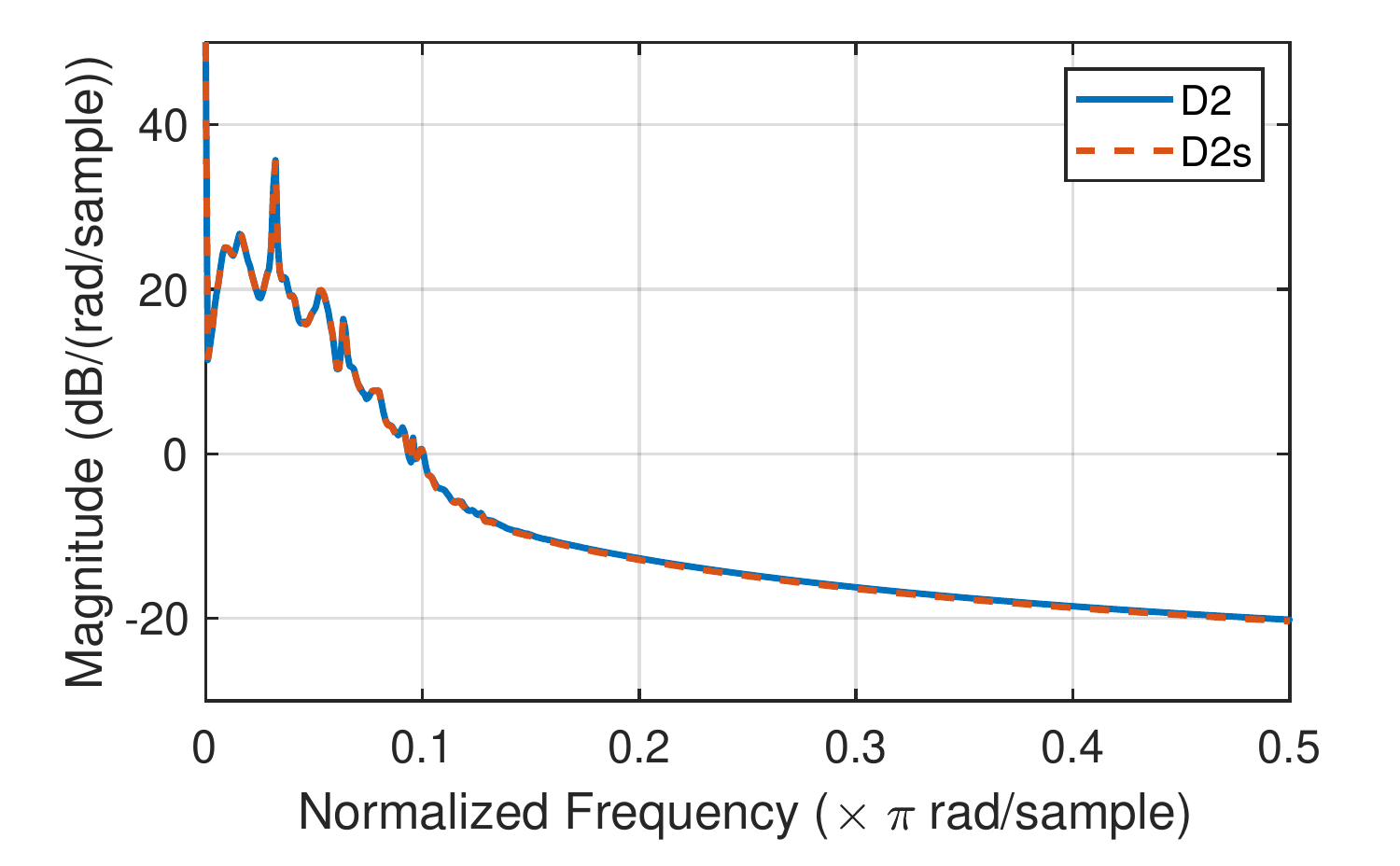}
    \caption{Estimated power spectral densities indicate similar frequency amplitude content for the generated chaotic loading and their surrogates. Upper left: first coordinate solution from Duffing's oscillator; 
    lower left: the stochastic surrogate signal to the first coordinate solution from Duffing's oscillator (which forms a chaotic-surrogate-pair along with the signal in the upper left figure); upper right: PSD of the first coordinate solution from Duffing's oscillator and its surrogate (D1 chaotic-surrogate pair); lower right: PSD of the second coordinate solution from Duffing's oscillator and its surrogate (D2 chaotic-surrogate pair).}
    \label{fig:spectra_loading}
\end{figure}

The surrogate data generation ensures that the surrogate random samples have nearly identical stress-level PDF and PSD, e.g., see~\figtwo{fig:PDF_loading}{fig:spectra_loading}.
The left two subplots in \fig{fig:spectra_loading} reflect the temporal dynamics of the time series phase portraits obtained by the delay-time embedding~\cite{kantz2004nonlinear}.
The upper-right plot depicts the phase portrait representation of the D2 time series, which indicates a low-dimensional determinism or manifold,
while the lower-left plot shows a cloud of curves where the deterministic temporal structure is no longer present. 
The linear auto-correlation functions of the chaotic loading and its surrogate are nearly identical despite the differences in their phase space representations.
The statistics of the load time histories, which include the mean, the root mean square (RMS) value, the standard deviation $\mathrm{Std(\cdot)}$, and the higher-order central moments (i.e., skewness and kurtosis) are calculated and summarized in Table~\ref{tab:statistics}. 
All statistics in Table~\ref{tab:statistics} indicate that the surrogate stochastic data possess identical statistics compared to their original chaotic counterparts.
Note that the generated loads are stationary and include around $10^4$ cycle ranges per cycle counting methods to mimic the prior experiments.

The crack growth simulations are run by repeatedly applying the generated loading, making the far-field stress periodic. 
Thus, the running crack size naturally introduces the nonstationarity in FCP calculation.

\subsection{Fatigue Crack Growth Simulation}
In the prior study~\cite{nguyen2013fatigue}, the authors pointed out that the linear cumulative damage rule overestimated the damage to the structure done by chaotic loading.
There are numerous of existing fatigue crack propagation models which consider load interaction effects based on various crack closure and opening hypotheses~\cite{elber1971significance,wheeler1972spectrum, willenborg1971crack,harter1999afgrow,huang2005fatigue,lu2010small, liu2017time}. 
These models capture the load interaction dynamics based on the adjustments to the crack opening stress level~\cite{newman1984crack,newman1992fastran}.
Among these methods, the \emph{crack closure model} is considered here for its algorithmic simplicity and its capability of considering load sequence effects.
Experimental observations that the crack is closed when the applied load is at zero, and the crack only propagates when the applied load is over the \emph{crack opening stress}, $\sigma_\mathrm{op}$, are the basis for the crack closure model.
The closure effect is characterized by the closure factor, $C_\mathrm{f}$, defined as the ratio of the crack opening stress to the maximum applied stress; and it can be determined from an empirical relationship,
\begin{equation}\label{eq:Cfeqn}
    C_\mathrm{f} = 1 - \left(\left(1-C_\mathrm{f0}\right)\left(1+0.6R\right)\left(1-R\right)\right),
\end{equation}
where $C_\mathrm{f0}$, the initial crack closure factor, is the only parameter needed for adjusting the closure effect for a given material.
For an accurate crack propagation simulation, this parameter needs verification according to fatigue test data.
In this paper, the material and structural properties are assumed to be identical. The temporal difference in the chaotic-surrogate pairs is also considered the sole contributor to fatigue life's observed variability.
Therefore,  $C_\mathrm{f0}$ is not meticulously selected based on material and fatigue testing database as it would be the same for the chaotic-surrogate pairs.
Before the simulation, we convert the original chaotic and the corresponding stochastic surrogate time series to the sequences of local load reversals (i.e., peak finding algorithm)~\cite{astm2005standard}.
A cycle in CAL is defined as the load variation from the minimum to the maximum and then to the minimum load~\cite{astm2005standard}.
However, there is no standard definition of a cycle under VA loading.
In this paper, to keep a consistent definition of a cycle, the sequence always starts with minimum stress during the conversion from signal to load reversals.
Therefore, we define a cycle as the local minimum stress followed by the local maximum stress, followed by the next local minimum stress.
During the simulation, the stress ratio, $ R $, is first determined by the first cycle's stress intensity factor ratio.
Then, the equivalent stress ratio for the given $C_\mathrm{f}$ is determined from \eq{eq:Cfeqn}.
One can refer to appendix A.2 for a detailed explanation of the calculation.

\begin{table}[]
\centering
\begin{tabular}{@{}lccclcccl@{}}
\toprule
      & \multicolumn{3}{c}{Experiments} &  & \multicolumn{3}{c}{Simulations} &  \\ \cmidrule(r){1-4} \cmidrule(l){6-9} 
      & Chaotic   & Surrogate   & Ratio  &  & Chaotic   & Surrogate  & Ratio  &  \\ \cmidrule(lr){2-4} \cmidrule(lr){6-8}
Exp-1 & 586       & 291         & 2.01   &  & 553       & 328        & 1.63   &  \\
Exp-2 & 432       & 266         & 1.62   &  & 384       & 374        & 1.02   &  \\
Exp-3 & 602       & 244         & 2.47   &  & 584       & 281        & 2.07   &  \\
Exp-4 & 607       & 307         & 1.98   &  & 582       & 282        & 2.06   &  \\
Exp-5 & 770       & 356         & 2.16   &  & 762       & 367        & 2.07   &  \\ \midrule
\multicolumn{9}{l}{CTFs are illustrated in $\times 10^3$ cycles }  \\ \bottomrule
\end{tabular}
\caption{Comparison of the CTFs between experiments and simulations for numerical model validation. Left group contains the CTFs using the rainflow-counting method to the experimentally obtained acceleration data; the corresponding ratios between chaotic case and their surrogates are given in the ratio column. Right group contains CTF statistics from simulations based on the same acceleration data.}\label{tab:CTF_comp}
\end{table}

\begin{figure}
    \centering
    \includegraphics[width = \textwidth]{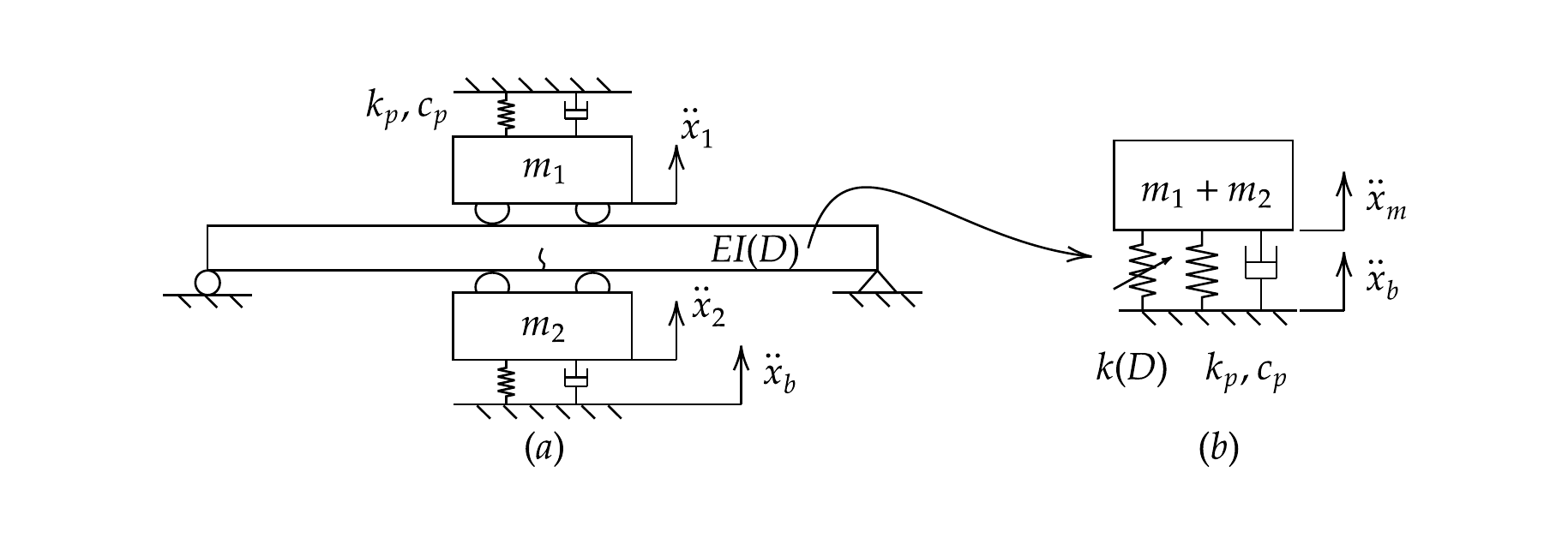}
    \caption{Schematics of the simulated fatigue experiment apparatus and its simplified one-degree-of-freedom (1-DOF) representative; (a) schematics of the fatigue experiment apparatus which uses a simply-supported aluminum beam specimen with damage (modeled as damage dependent flexural rigidity $EI(D)$), driven by base acceleration $\ddot{x}_b$, and constrained by pneumatic supporting system, $k_\mathrm{p}$, $c_\mathrm{p}$, and $m_{1,2}$; (b) a 1-DOF representative which models the cracked beam as a nonlinear spring $k(D)$ and the damage is driven by the relative acceleration $\ddot{x}_\mathrm{rel} = \ddot{x}_\mathrm{m} - \ddot{x}_\mathrm{b}$. Note that the fixed boundaries are connected as one rigid body.}
    \label{fig:schematics}
\end{figure}

We use AFGROW 5.3 environment~\cite{afgrownet}---which has a variety of fatigue crack propagation models that consider load interaction effects---for the simulations.
This package also provides a crack propagation rate database for physics-based simulations.
The structure's material under consideration is aluminum alloy 2024-T3, and the corresponding material properties and model parameters can be obtained from appropriate references Refs.~\cite{ray2001fatigue1, ray2001fatigue2, newman1992fastran,afgrownet}. 
Here, we use the one from the Fracture Mechanics Database organized by LexTech, Inc., for its consistency and stress ratio dependence~\cite{afgrownet}.
We simulate a three-point-bending specimen with a single-edged-through crack in AFGROW 5.3 to validate the model's predictive ability. 
This simulated model is based on previous experimental studies~\cite{nguyen2013fatigue,li2019variable}; where the geometry is kept identical, see \fig{fig:schematics}.
The initial crack length is set to $1.27\times10^{-3}$\,m and the final crack length is $6\times10^{-3}$\,m.
Then, the simulation uses the corresponding experimentally obtained relative acceleration $a_\mathrm{rel}$ data as the input load, see subplot (b) in \fig{fig:schematics}.
Table~\ref{tab:CTF_comp} illustrates the comparison of simulated and experimentally obtained CTFs, which shows a good correlation in the CTFs and CTF ratios between the simulation and experiment (erroneously recorded acceleration data during the experiment is the cause of the exception in test-2).
The experimental strain measurement under CAL using a healthy beam specimen and accounting for a stress concentration factor for notched specimens~\cite{pilkey2008peterson} guided the normalization of all the load reversals with an identical mean load level of 25\,MPa and standard deviation of 70\,MPa.
Based on the simulation results, we assume that the model is reliable for future life estimation using the generated irregular loading.  

\section{Proposed Fatigue Life Estimation Framework}
The proposed framework is a procedure which has three main components: (1) the overload identification from load reversals, (2) the modeling of the load sequence effects, and (3) the estimation of irregularity for a given loading to ensure improvement to the fatigue life estimation using CDR.

\subsection{Rainflow-Counting-Based Overload Identification}
A key feature to the consideration of the crack growth retardation effect is identifying the overloads in a given load time-history.
According to many experimental studies~\cite{potter1971effect,bathias1978mechanisms}, the overload ratio (or the overload amplitude) and the normalized period (or, equivalently, the rate of occurrence of overload) govern the crack retardation dynamics.
These experiments mostly focused on programmed periodic overloads or periodic sequences with less irregularity, where the definition and identification of overloads are simple.
However, whenever the given loading is highly irregular, the definition of overload becomes less clear.
The indices of load reversals identified in the rainflow-counting algorithm can be extracted and classified as candidates of overloads, which allows us to identify the overload consistently while retaining as much physical sense of overload as possible.
To initiate the identification, one has to obtain the rainflow-counting matrix for a given load time-history.
Refs.~\cite{astm2005standard,amzallag1994standardization,mathworks} provide a detailed explanation of the rainflow-counting algorithm.
The rainflow-counting method accounts for the hysteresis effects of the stress-strain response of a structure, and it is also the cycle counting method that accounts for the ``memory effect.''
During the rainflow-counting procedure, first, the loading excursion towards the local maximum stress point is counted as a half-cycle. 
Then, the unloading excursion towards the local minimum stress counts as another half-cycle.
Then, the cycles in between the two identified half-cycles are saved for the second round of counting, during which we identify the smaller excursions of successive loading and unloading cycle ranges as full cycles.
This process goes through all the reversals for a given load time history until we identify all successive loading and unloading excursions.
Finally, if there are any leftover small local cycles, we count them as full cycles and store their corresponding starting and ending indices.
Here, we assume that the identified loading half-cycles form \emph{mesoscale} plastic zones at the crack front.
On the other hand, the unloading half-cycles provide the onset of retarded cycles after the mesoscale maximum plastic zones.
Similarly, full cycles, identified as the successive loading and unloading cycles, also provide the location information of \emph{local} overload and the local onset of retarded cycles.
Based on this discussion, we classify overloads into four class types:
\renewcommand{\labelenumi}{\Alph{enumi}}
\begin{enumerate}
    \item[A:] happens at the ending indices of the loading half cycles. This point gives the reversal index that forms a mesoscale overload plastic zone.
    \item[B:] happens at the onset indices of the unloading half cycles. This type's index indicates the following cycles are retarded with the most probability since this onset is the maximum stress for a given load time-history.
    \item[C:] happens at overload indices corresponding to the ending of local successive loading full cycles.
    \item[D:] happens at overload indices corresponding to the onset of local successive unloading full cycles.
\end{enumerate}
The overload classes of type A and type B (or type C and type D) are, most of the time, mutually exclusive.
If A and B (or, C and D) are connected without any in-between reversals, the overload identified is duplicated.
Only the unique indices are selected after the identification concludes to eliminate duplication.
Unlike the conventional sorting from the rainflow algorithms, which sort the half-reversals in front of the rest of the smaller full cycles, we sort the identified rainflow overloads according to their indices of occurrence identical to the original loading.
This sorting facilitates the imposition of load sequence effect to the rest of the affected cycles, see Table \ref{tab:rainflowalg} for more detailed descriptions on the overload identification algorithm.

\begin{table}[]
\centering
\begin{tabular}{@{}llllllllllll@{}}
\toprule
\multicolumn{12}{l}{Algorithm for rainflow-counting Based Overload Identification} \\ \midrule
\bfseries{Require:} &
  \multicolumn{11}{l}{\cellcolor[HTML]{EFEFEF}Given rainflow matrix, $M_\mathrm{rf}$, from rainflow-counting algorithm} \\
 &
  \multicolumn{11}{l}{\textbf{while} index of reversal, i, is not depleted \textbf{do}} \\
 &
  \multicolumn{11}{l}{\cellcolor[HTML]{EFEFEF}Sort $M_\mathrm{rf}$ according to the onset index of identified reversals} \\
 &
  \multicolumn{11}{l}{\begin{tabular}[c]{@{}l@{}}Identify overload Type-A from reversals which are counted as half cycles\\ whose starting index is odd, and return a logical array $\mathbf{A}$.\end{tabular}} \\
 &
  \multicolumn{11}{l}{\cellcolor[HTML]{EFEFEF}\begin{tabular}[c]{@{}l@{}}Identify overload Type-B from reversals which are counted as half cycles\\ whose starting index is even, and return a logical array $\mathbf{B}$.\end{tabular}} \\
 &
  \multicolumn{11}{l}{\begin{tabular}[c]{@{}l@{}}Find overload Type-C from reversals which are counted as one full-cycle whose\\ starting index is odd.\\ Identify overload Type-C if the mean-stress-corrected stress amplitude is larger\\
  than its next one,\\ $\sigma_\mathrm{ar}^{(i)}> \sigma_\mathrm{ar}^{(i+1)}$, and return a logical array $\mathbf{C}$.\end{tabular}} \\
 &
  \multicolumn{11}{l}{\cellcolor[HTML]{EFEFEF}\begin{tabular}[c]{@{}l@{}}Find overload Type-D from reversals which are counted as one full-cycle whose\\ starting index is even.\\ Identify overload Type-D if the mean-stress-corrected stress amplitude is larger\\
  than its previous one,\\ $\sigma_\mathrm{ar}^{(i)}>\sigma_\mathrm{ar}^{(i-1)}$, and return a logical array $\mathbf{D}$.\end{tabular}} \\
\bfseries{Return:} &
  \multicolumn{11}{l}{Logical array $\mathbf{OL} = \mathrm{unique}([\mathbf{A}\, \mathbf{B}\, \mathbf{C}\, \mathbf{D}]$)} \\ \bottomrule
\end{tabular}
\caption{Algorithm for rainflow-counting based overload identification.}
\label{tab:rainflowalg}
\end{table}

\subsection{Modeling of Load Sequence Effects}

After identifying all the overloads, we still need to know how these overloads affect structural degradation dynamics.
We answer this question by assuming the crack growth retardation effect as the most notable load sequence effect and assuming the PDF of the retarded $\sigma_\mathrm{rar}$ drives the dynamics of the damage accumulation. 
This can be seen as modifying the probability density function $p(\sigma_{a})$ in \eq{eq:ettf} to the mean-stress-corrected stress amplitude, $\sigma_\mathrm{ar}$.
The value of $\sigma_\mathrm{ar}$ is calculated using the Walker equation $\sigma_\mathrm{ar} = (\sigma_\mathrm{max}\sigma_\mathrm{m})^\gamma$ with the mean stress contribution coefficient $\gamma = 0.5$~\cite{walker1970effect,smith1970stress,dowling2009mean}.
Then, we consider the retardation effect as a temporal weight to $\sigma_\mathrm{ar}$, where $\sigma_\mathrm{ar}$  is further adjusted by a retarded value of stress (e.g., the residual stress variation), $\Delta\sigma_\mathrm{R}$.
We call it the retarded mean-stress-corrected stress amplitude, $\sigma_\mathrm{rar}$, and express it as,
\begin{equation}\label{eq:sigmarar}
    \sigma_\mathrm{rar} = \sigma_\mathrm{ar} + \Delta\sigma_\mathrm{R}.
\end{equation}
Therefore, characterization of the overload retardation effect---the residual stress variation---from the identified overloads is required.
\subsubsection{Earlier studies on modeling load sequence effects}
Numerous studies investigated this effect by correlating the affected life span of structures to the rate of occurrence of overloads and the magnitudes of the overloads~\cite{bathias1978mechanisms,potter1971effect,skorupa1998load,skorupa1999load,borrego2003evaluation,zhang2012situ}.
Researches reported that the strength of retardation effects are related to the {\em overload ratio} (OLR), defined by $\alpha^{(i)} = \sigma_\mathrm{max}^{(i)}/\sigma_\mathrm{max}^{(i+1)}$~\cite{bathias1978mechanisms}; and arrested crack happens when this value exceeds certain threshold~\cite{geary1992review}.
A more comprehensive experimental study is by J. M. Potter~\cite{potter1971effect} from which it suggests that \emph{the relative fatigue life is a function of the relative overload period and the overload amplitude}.
The relative fatigue life is the overload affected fatigue life divided by the unaffected fatigue life, $N_\mathrm{f}/N_{\mathrm{f^{'}}}$.
Furthermore, the relative overload period is the number of cycles between overloads divided by the unaffected fatigue life, $N_{\mathrm{ol}}/N_{\mathrm{f^{'}}}$.
The relative fatigue life increases to a maximum value when the applied overload period is about 5 percent to 10 percent of fatigue life without the presence of overloads, and it drops asymptotically to the relative fatigue life of about 1.5 as the overload period increases.
In contrast, when the overloads occur too frequently, the relative fatigue life drops to one-fourth of the no-overload case.
The dynamic residual stress, $\sigma_\mathrm{R_{d}}$, is expressed as an exponential decay function with an initial value of a deviation from the equilibrium residual stress; this deviation disappears when this dynamic residual stress decays to 10 percent of its initial value~\cite{potter1971effect}:
\begin{equation}
    \sigma_\mathrm{R_d} = \left(\sigma_\mathrm{R}-\sigma_{\mathrm{R}_\mathrm{eq}}\right)\mathrm{exp}\left[\left(N/N_\mathrm{eq}\mathrm{log}(0.1)\right)\right]
\end{equation}
The difference between the total residual stress, $\sigma_{\mathrm{R}_\mathrm{d}}$, and the \emph{equilibrium residual stress}, $\sigma_{\mathrm{R}_\mathrm{eq}}$, is the amplitude of the transient residual stress immediately after the overload cycle; the mitigation of the dynamic residual stress is governed by the exponentially decaying function; $\mathrm{log}(0.1)/N_\mathrm{eq}$ is the decay rate determined from the number of cycles needed to cause the dynamic residual stress to return to $10\%$ of its initial value upon the occurrence of an overload.
The decay time is described by the equilibrium period,
\begin{equation}\label{eq:Neq}
N_\mathrm{eq} = A/[(\sigma_\mathrm{max})^p(\sigma_\mathrm{m})^q],
\end{equation}
which is a function of proportionality constant $A$; the maximum stress $\sigma_\mathrm{max}$; the mean stress $\sigma_\mathrm{m}$; and their corresponding contribution exponents, $p$ and $q$.
Further, the stress amplitude is affected by this dynamic residual stress, and it can be expressed as
$
   \sigma_r = \sigma + \sigma_{\mathrm{R}_\mathrm{eq}} -  \sigma_\mathrm{R_d}
$,
where $\sigma_r$ is the resultant stress after considering the dynamic residual stress variation ($\Delta\sigma_{\mathrm{R_d}} = \sigma_\mathrm{R_{eq}} - \sigma_\mathrm{R_d}$) as the deviation from the equilibrium residual stress $\sigma_{\mathrm{R}_\mathrm{eq}}$.
However, the equilibrium period in \eq{eq:Neq} does not have a proper dimension of a cycle.
\subsubsection{The proposed load sequence effect model}
\begin{figure}[htbp]
    \centering
    \includegraphics[width = 0.45\textwidth]{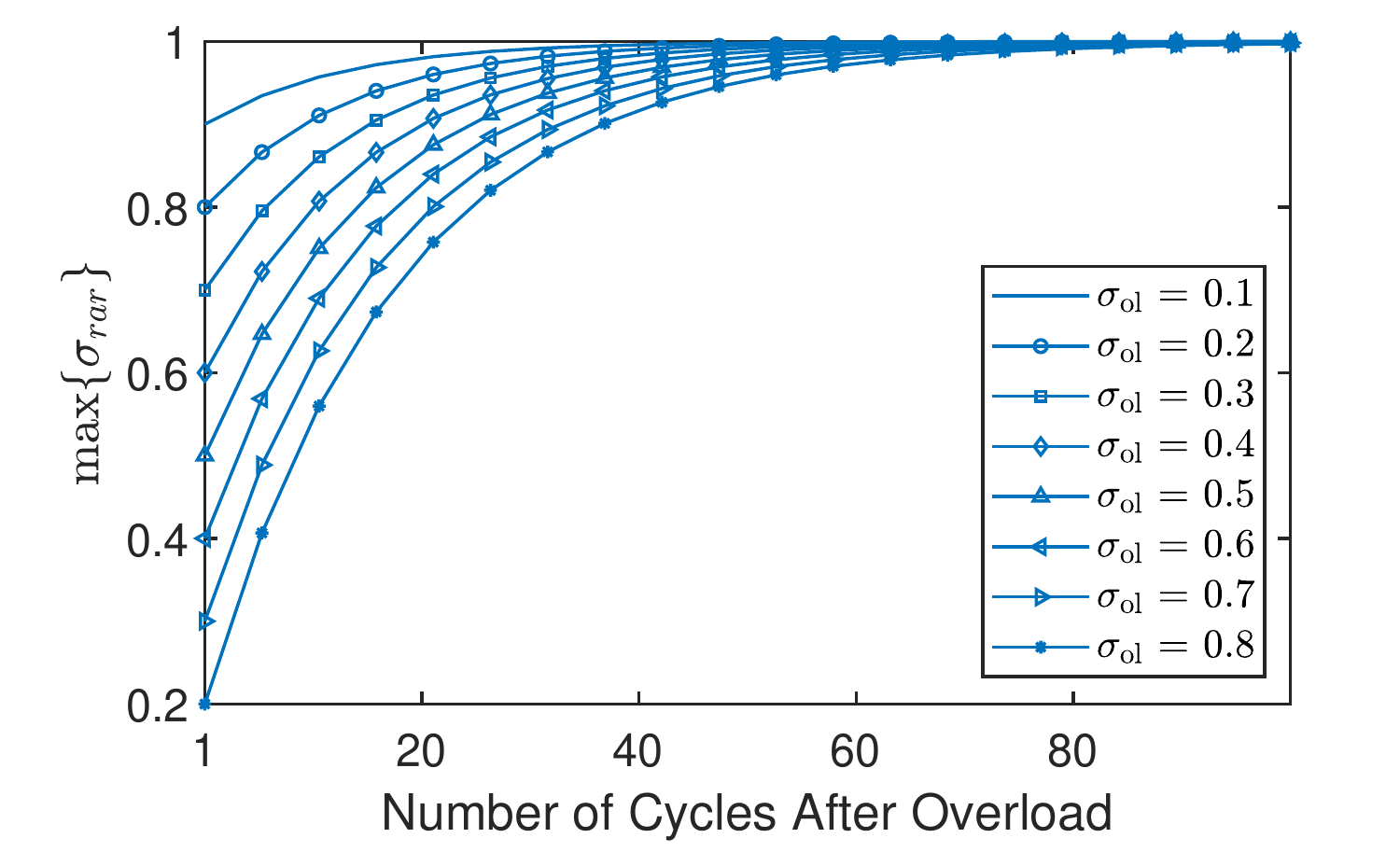}
    \includegraphics[width = 0.45\textwidth]{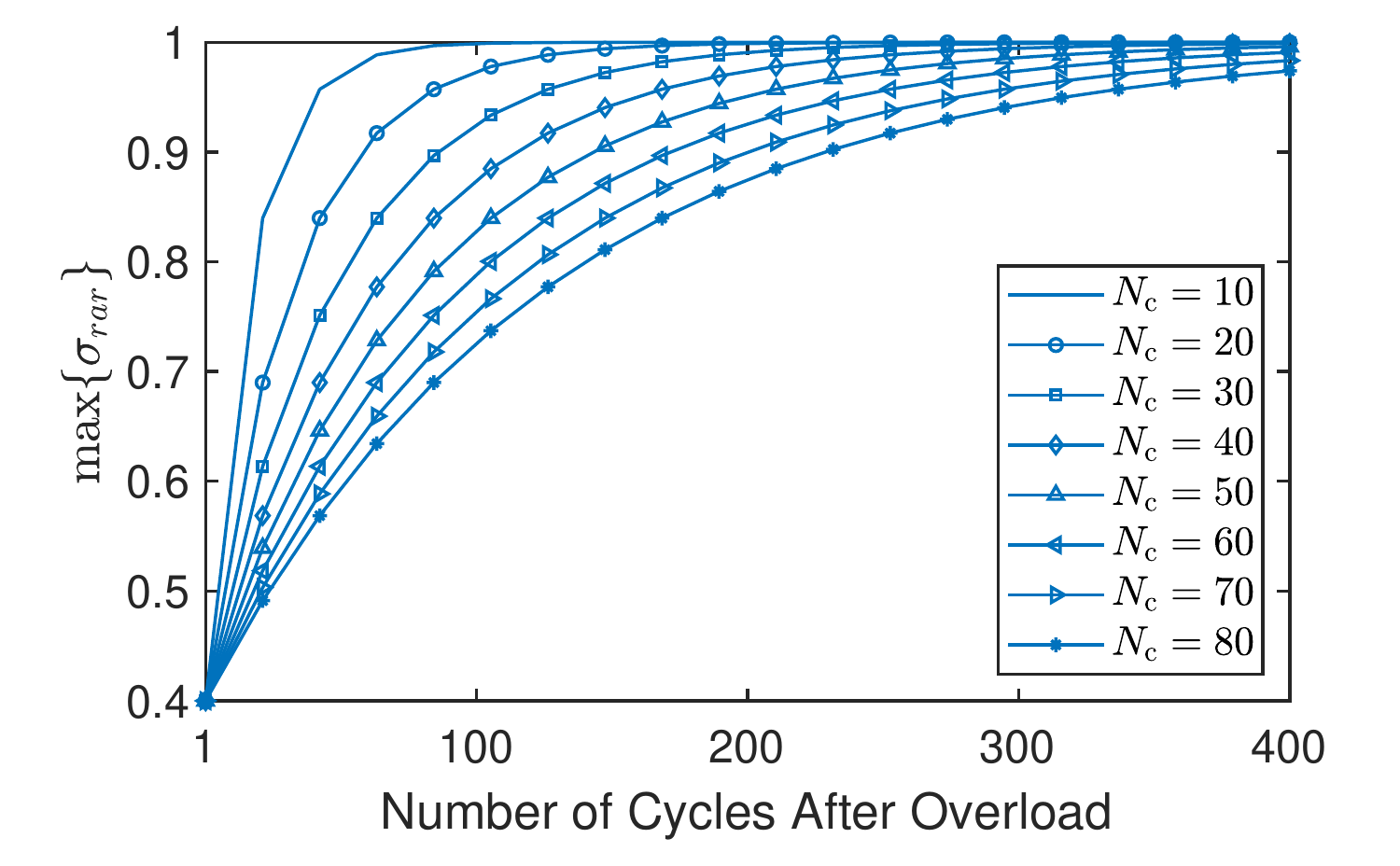}
    \caption{Maximum stress profile of retarded stress history from an imaginary harmonic loading with normalized magnitude. Left: maximum stress profile of the adjusted load time history with $N_\mathrm{c} = 10$ and $\sigma_{ol}$ varies from $0.1 - 0.8$; right: maximum stress profile of the adjusted load time history with $\sigma_{ol} = 0.6$ and $N_\mathrm{c}$  varies from $10 - 80$.}
    \label{fig:t_weight}
\end{figure}

\begin{figure}[htbp]
    \centering
    \includegraphics[width = 0.45\textwidth]{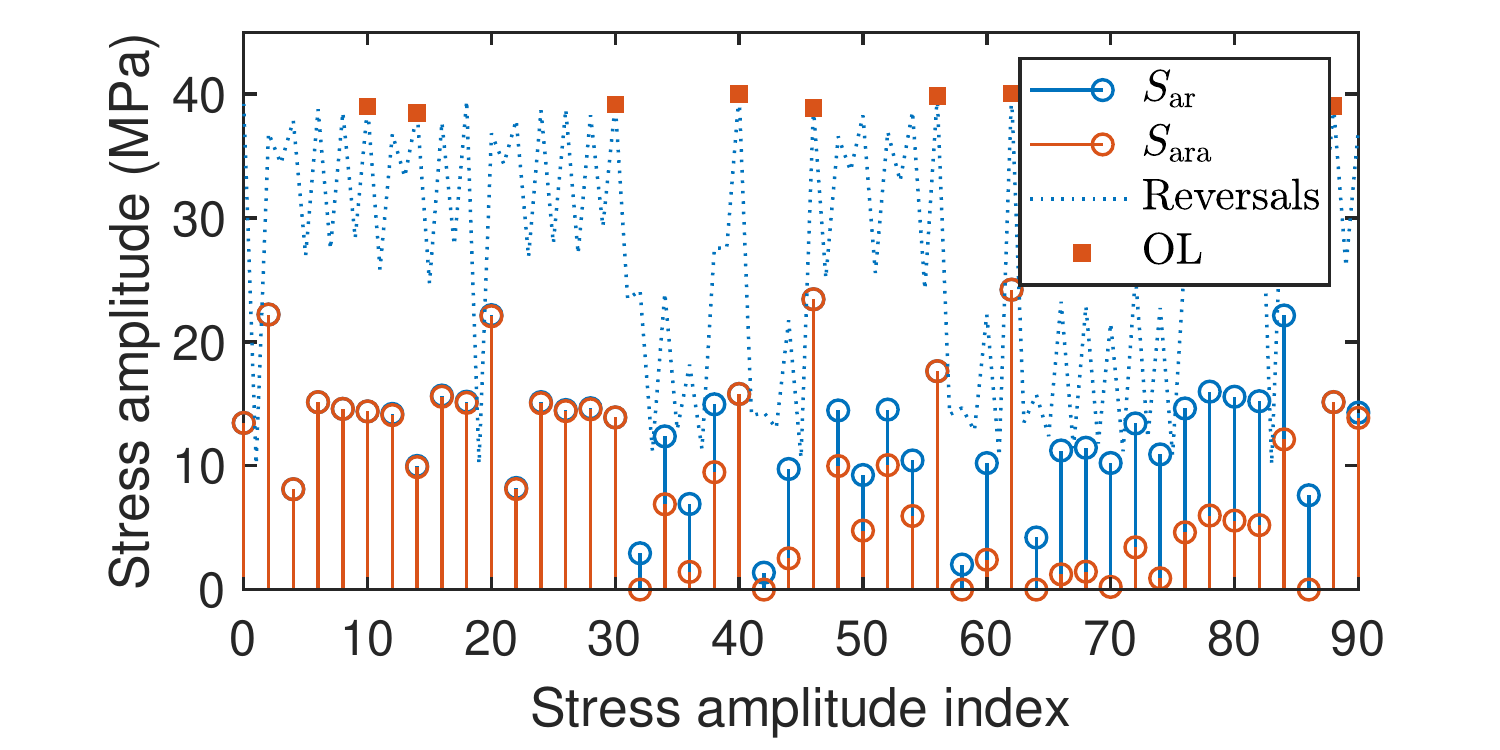}
    \includegraphics[width = 0.45\textwidth]{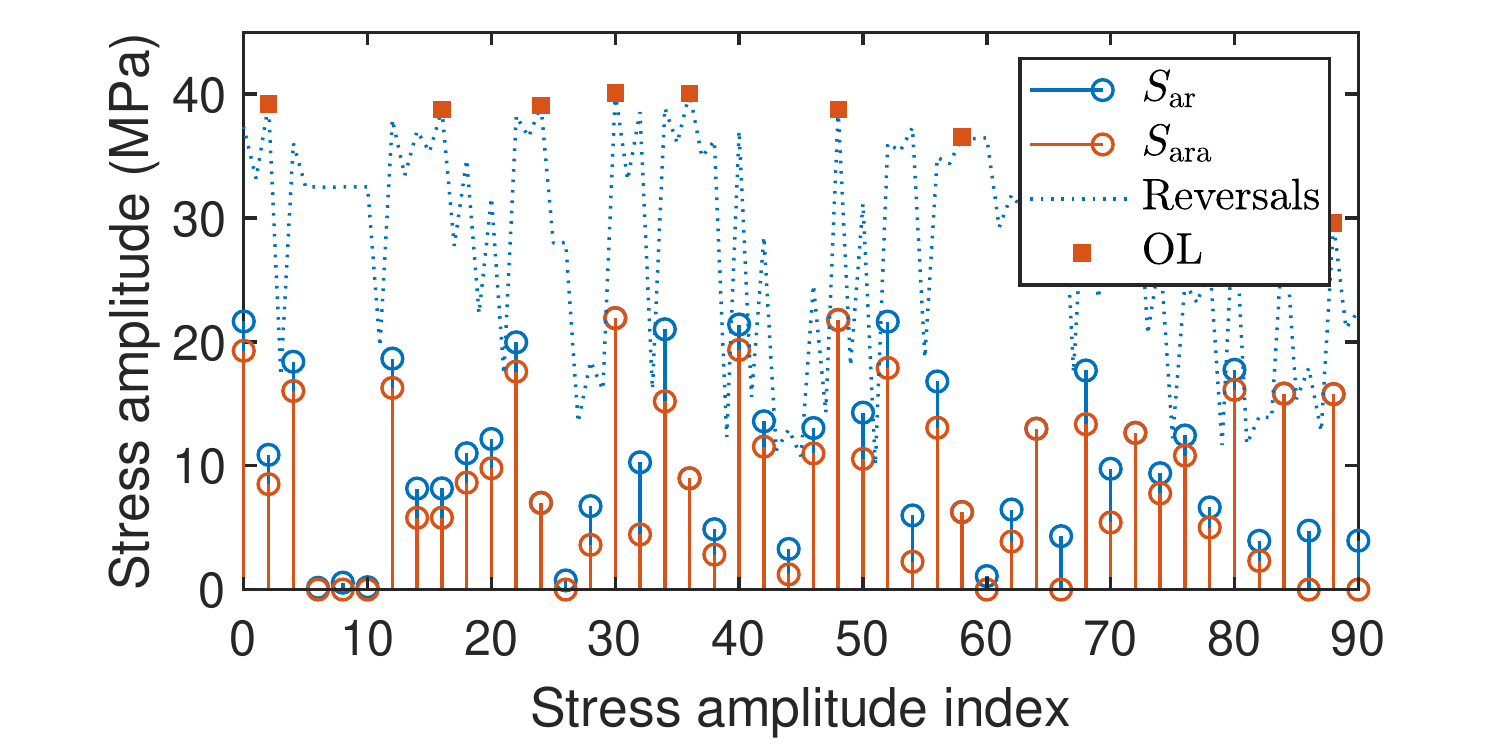}\\
    \includegraphics[width = 0.45\textwidth]{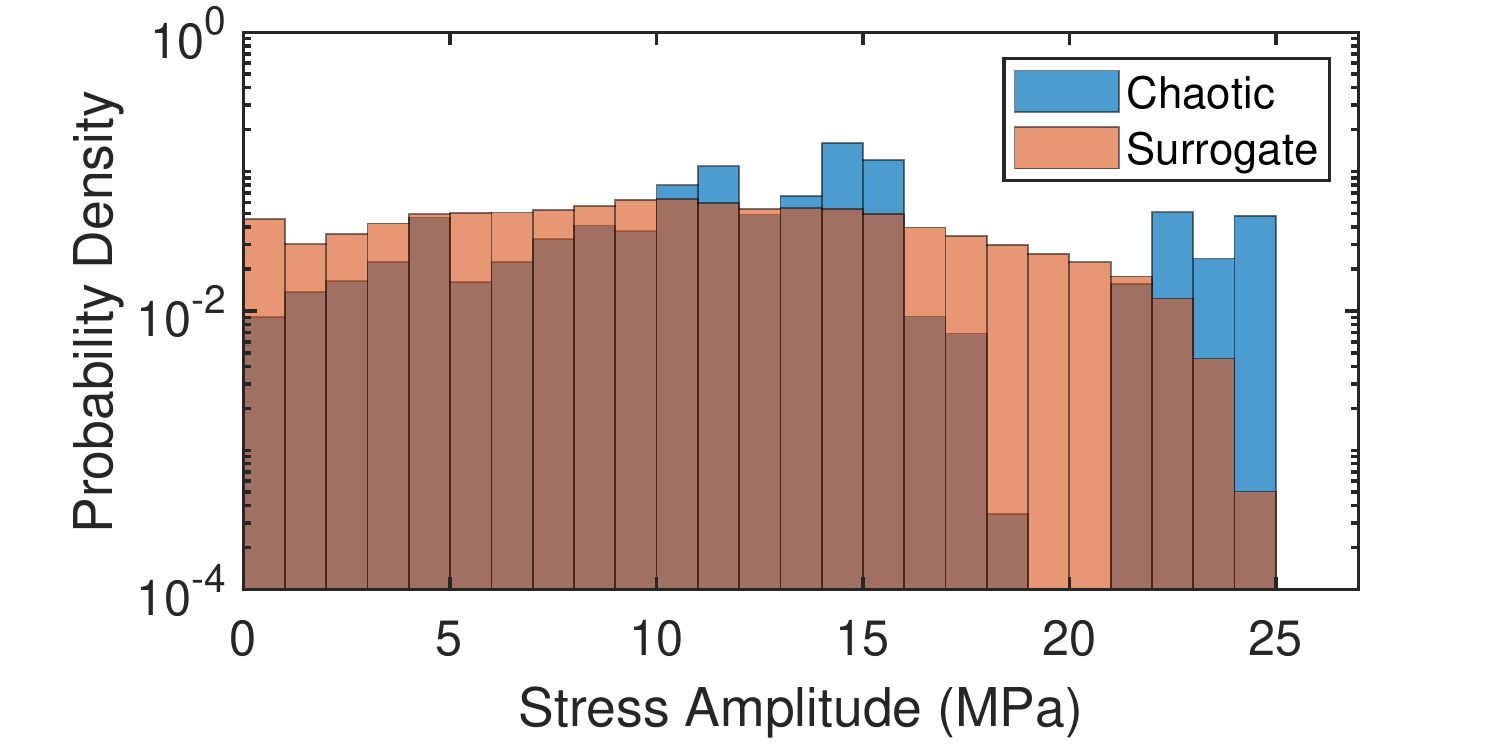}
    \includegraphics[width = 0.45\textwidth]{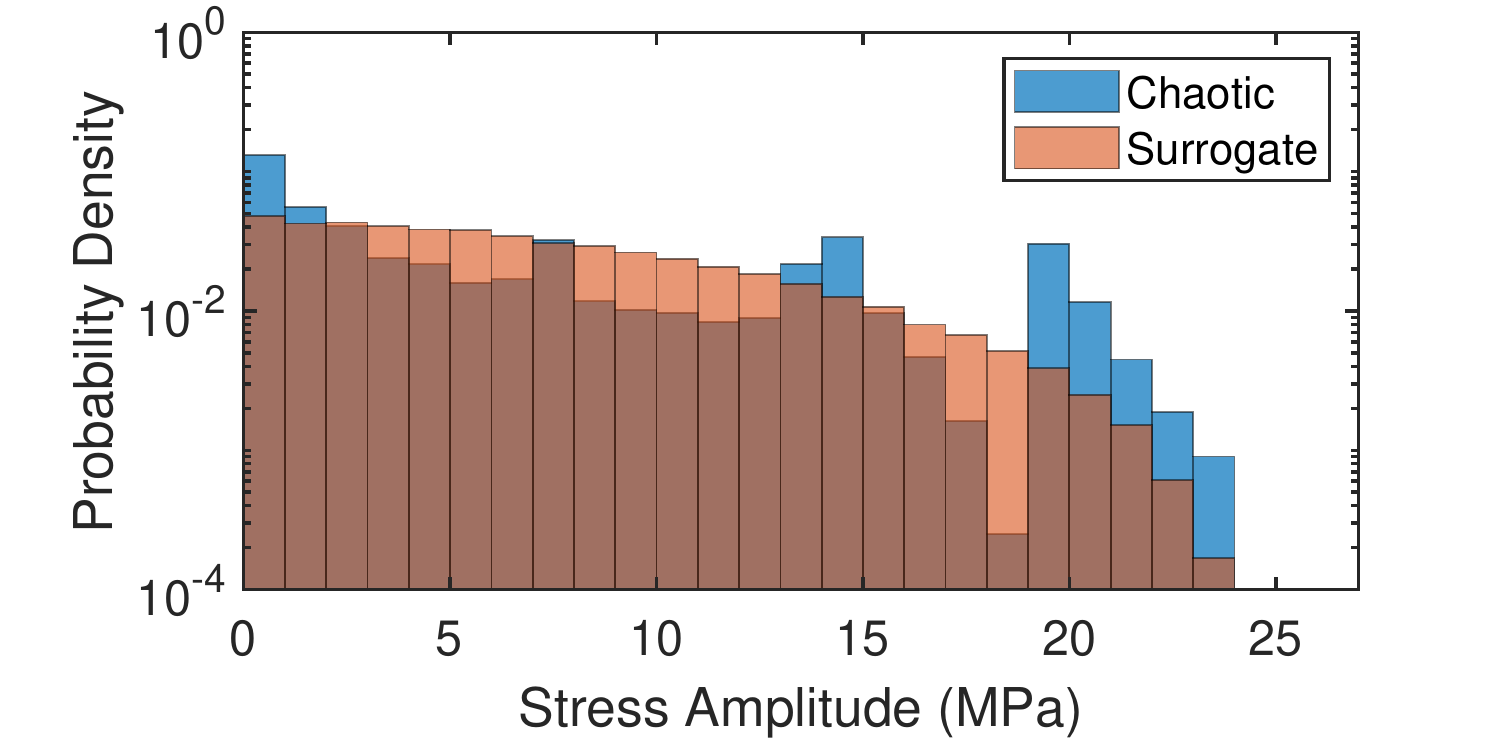}
    \caption{Upper: Combined view of rainflow based overload identification and temporal retardation; lower: estimated mean stress adjusted stress amplitude PDF. Upper-left: D1 synthetic loading; upper-right: D1s synthetic case. Lower-left: comparison of stress amplitude PDF before retardation adjustment; lower-right: comparison of stress amplitude PDF after retardation adjustment.}
    \label{fig:CR-PLD}
\end{figure}

Based on Potter's formulation, we propose a load sequence effect model that corrects the overload period's dimensionality and replaces the residual stress variation by the modified dynamic residual stress to obtain $\sigma_\mathrm{rar}$ without explicitly tracking the residual stress for every cycle.
Compared to \eq{eq:Neq}, we propose a new description of an equilibrium period with the dimension of a cycle, using OLR as the non-dimensional governing factor.
The equilibrium period can be expressed as,
\begin{equation}\label{eq:Neq1}
    N_\mathrm{eq}^* = N_{c}\times\mathrm{exp}(\alpha)
\end{equation}
where $N_\mathrm{c}$ is proportionality constant that has the dimension of cycle.
When the applied load is irregular, the definition and determination of the equilibrium residual stress become tricky, and the overall local stress fluctuation becomes highly dynamic.
Therefore, we propose a dynamic residual stress formulation as follows,
\begin{equation}\label{eq:sigmadnew}
    \sigma_\mathrm{R_{d}}^* = \rho_\mathrm{ol} \sigma_\mathrm{ol}\,\mathrm{exp}\left[N/N_\mathrm{eq}^*\mathrm{log}(r)\right],
\end{equation}
where $\rho_\mathrm{ol}$ is the overload amplitude ratio that governs how much initial residual stress is introduced when overload occurs; $\sigma_\mathrm{ol}$ define {\em the overload amplitude} as the difference between two consecutive peaks where the first peak is an overload. $\mathrm{log}(r)$ with $r\in\left(0, 1\right)$, as a generalization to $\mathrm{log}(0.1)$, is the fraction of dynamic equilibrium residual stress that determines the remaining residual stress variation when the retardation effect disappears.
\fig{fig:t_weight} shows the effect of the proposed temporal retardation weight, where the left plot shows the relationship between the maximum stress profile of the retarded load time history from an imaginary harmonic loading, with normalized amplitude between zero and one, with fixed $N_\mathrm{c}$ and varying $\sigma_\mathrm{ol}$. The right plot shows the effect of $N_\mathrm{c}$ with fixed $\sigma_\mathrm{ol}$.
We further assume the previously formed residual stress effects canceled immediately whenever a new overload occurs to reduce the computational complexity.
Therefore, the corresponding retarded mean-stress-corrected stress amplitude can be expressed as a function upon the occurrence of a given overload, $\sigma_\mathrm{ol}(j)$. 
By combining \eq{eq:sigmarar}, \eq{eq:Neq1}, and \eq{eq:sigmadnew}, we derived the explicit expression for the retarded mean-stress-corrected stress amplitude history,
\begin{equation}\label{sigmarar}
    \sigma_\mathrm{rar}{(j)} = \sigma_\mathrm{ar} - \sigma_\mathrm{R_d}^*(j) = \sigma_\mathrm{ar} - \rho_\mathrm{ol}\sigma_\mathrm{ol}\mathrm{exp}\left[\frac{\mathrm{log}(r)}{N_\mathrm{c}}\cdot\frac{N}{\mathrm{exp}(\alpha{(j)})}\right],
\end{equation}
in which $\rho_\mathrm{ol}$ and $\frac{\mathrm{log}(r)}{N_\mathrm{c}}$ are material/structural properties, $j$, being the index of overload, is determined from the proposed overload identification procedure, other variables are determined from the given load time histories.
In this way, \eq{sigmarar} can be applied as a temporal vectorial weighting function to the identified overload set, which is a subset of all the cycles in a given loading.
This formulation also allows one to obtain the load sequence effects without explicitly tracking the residual stress for every cycle in a cycle-by-cycle manner.
Since the overload occurrence in this study is frequent, the effect of $r$ on the overload retardation contribution is limited, and we keep $r = 0.1$ in the following results.
The retarded load time histories after adjustment are illustrated in \fig{fig:CR-PLD}, where a direct comparison between the original mean-stress-corrected stress amplitude and the one after adjustment can be observed after the identified overloads, using the proposed framework.
From the perspective of damage estimation, the temporal adjustment changes the PDF of stress amplitude metrics.
This changes in stress-amplitude statistics are also illustrated in \fig{fig:CR-PLD}, from which one can observe that the proposed retardation model modifies the stress-amplitude PDFs in such a way that the probability density of very high stress amplitudes decreases while its low--medium amplitude counterparts being amplified for the surrogate loading.
Moreover, the multi-mode distribution of the chaotic case changes to a more connected one, weighted towards the lower stress amplitudes that below $15\mathrm{MPa}$, see lower subplots in \fig{fig:CR-PLD}.

\subsection{Overload Rate Correction Factor}
\begin{figure}[htbp]
    \centering
    \includegraphics[width = 0.45\textwidth]{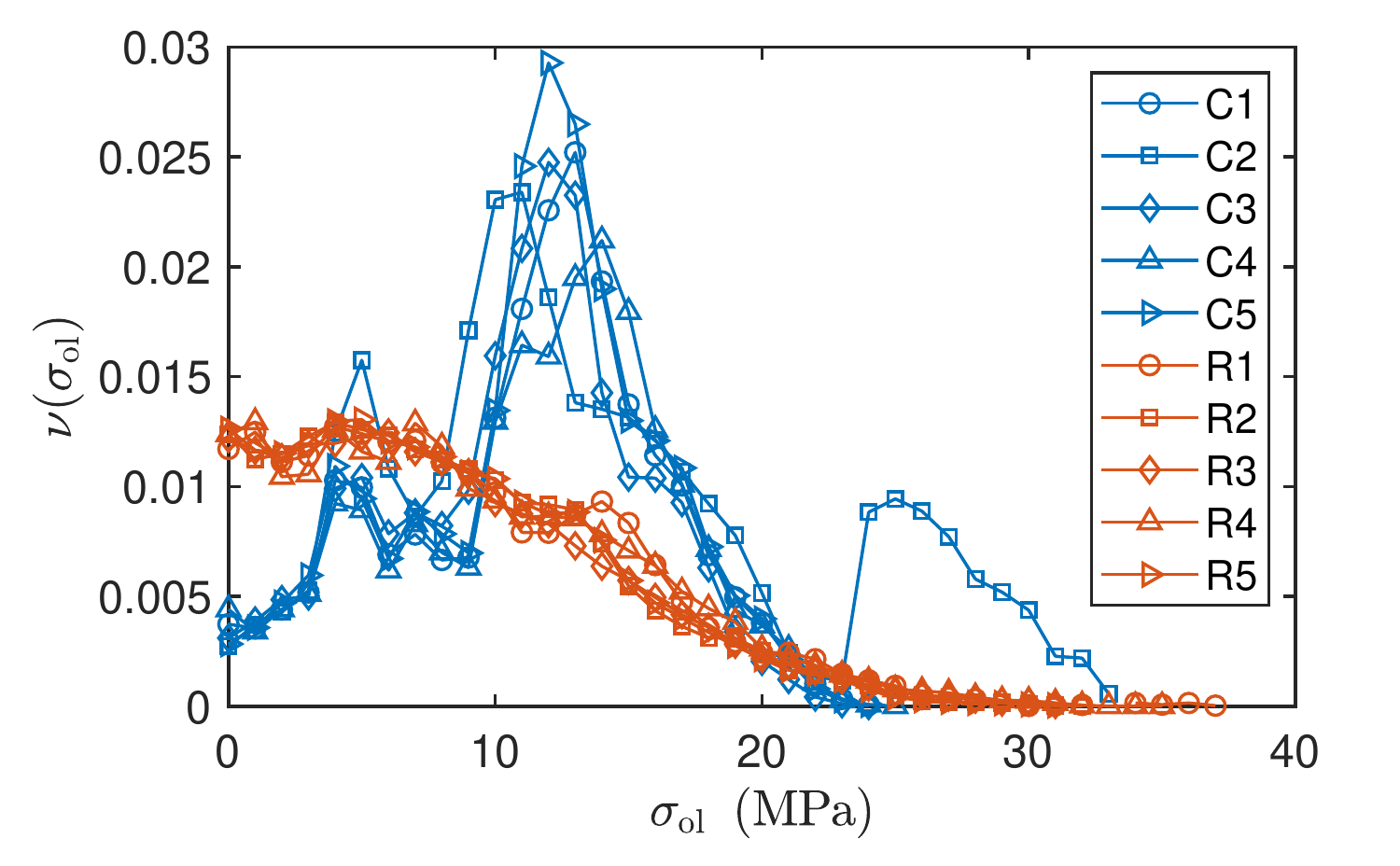}
    \includegraphics[width = 0.45\textwidth]{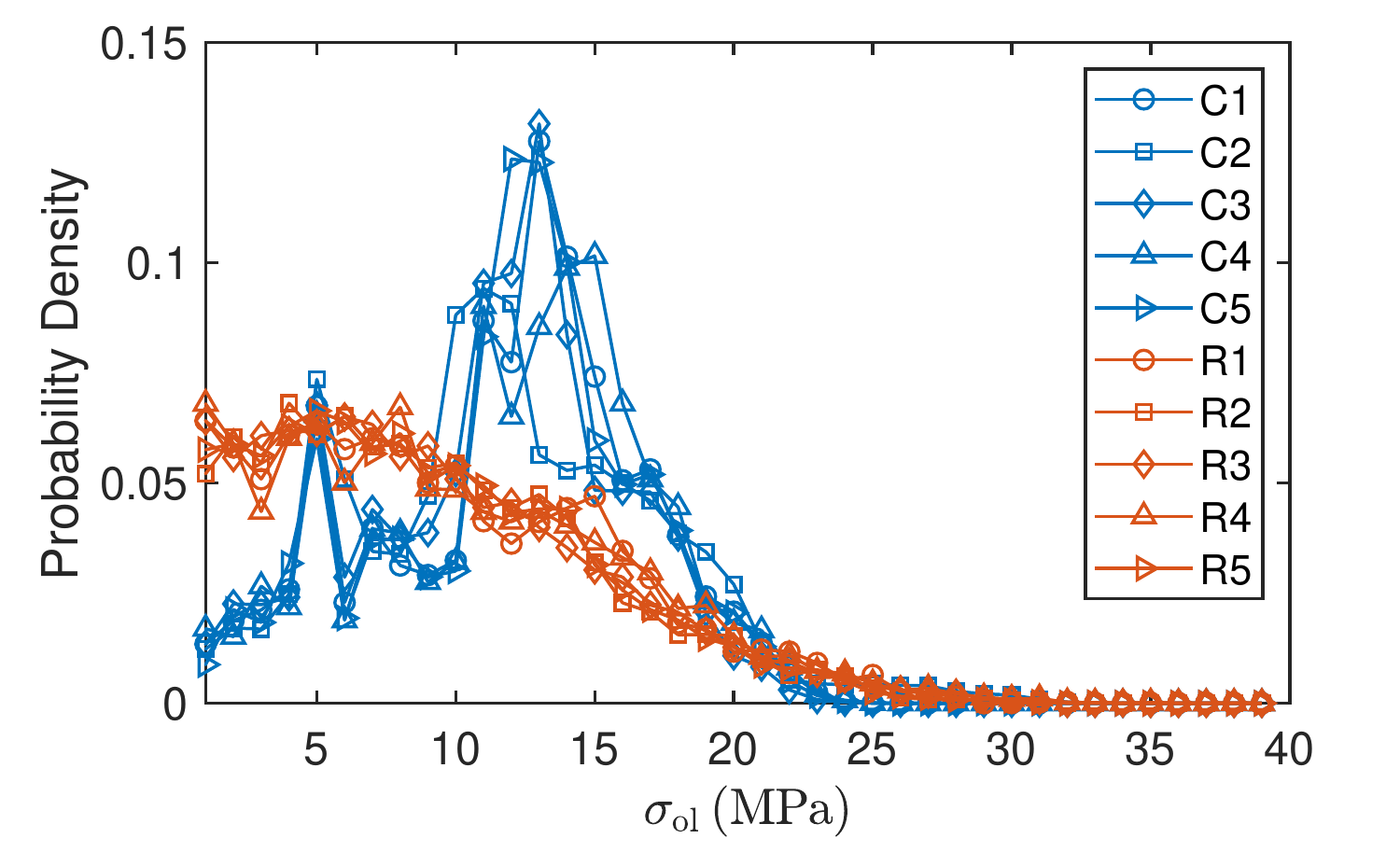}   
    \caption{Comparison of overload amplitudes statistics from experimental acceleration data. Left: expected rate of occurrence of overload amplitudes as a function of overload cycle ranges; right: PDF of overload amplitudes as a function of overload amplitude.}
    \label{fig:ERol_exp}
\end{figure}

One key step in fatigue life estimation using the CDR is estimating the applied load's occurrence rate. 
It works as an averaged load reversals occurrence rate, which scales the inverse of the estimated damage to a proper dimension of estimated life. 
For Gaussian random processes, the rate estimator is usually selected as the rate of occurrence of mean up crossing, $\nu^+$~\cite{lin1967probabilistic,benasciutti2005spectral}. 
In this paper, however, the loading is non-Gaussian and is not limited in the frequency bandwidth. 
Instead of using $\nu^+$, the rate of occurrence of peaks is selected, since it is not easy to define mean up-crossing for highly irregular non-Gaussian loading. 
To account for the loading rate for non-Gaussian loading, studies~\cite{marquis2011fatigue,braccesi2009frequency,bohm2020fatigue} indicate the use of irregularity factor to generalize the results of Gaussian random loading to accommodate the non-Gaussian cases through characterization of the PSD of the applied loading. 
However, since the considered loads in this study have nearly identical PSD and statistical distributions, such an irregularity factor cannot adjust the loading rate correctly. 
It is shown in the previous studies~\cite{potter1971effect,bathias1978mechanisms} that there exists a nonlinear relation between the relative fatigue life and the rate of occurrence of overloads.  
Furthermore, it is also intuitive to characterize the irregularity of a general loading based on the PDF of overload since the occurrence of overloads indicates irregular amplitude fluctuations in a given loading. 
Therefore, we propose to use the characteristics of the overload amplitudes as out rate correction factor. 
The characteristics of overload amplitude should provide two important information for life estimation considering the load retardation effect, the overload amplitudes, and their corresponding rate of occurrence. 
This correction should help adjust the overall loading rate, which provides proper scaling to the considered loading's estimated life.
One way to characterize the overload amplitude is to define the rate of occurrence of overload amplitudes (the inverse of the overload period) as a function of overload amplitude,
\begin{equation}
    \nu_\mathrm{ol}\left(\sigma_\mathrm{ol}^{(k)}\right) = \frac{1}{P_\mathrm{ol}\left(\sigma_\mathrm{ol}^{(k)}\right)} = \frac{1}{\frac{1}{M^{(k)}}\sum_{j=1}^{M^{(k)}}\left(I_{\sigma_\mathrm{ol}^{(k)}}(j) - I_{\sigma_\mathrm{ol}^{(k)}}(j-1)\right)}
\end{equation}
where, $P_\mathrm{ol}\left(\sigma_\mathrm{ol}^{(k)}\right)$ is the overload period as a function of overload amplitude and $k$ is the $k^\mathrm{th}$ discretized overload amplitude; $M^{(k)}$ is the number of occurrence of overload for the $k^\mathrm{th}$ discretized overload amplitude; $I(\cdot)$ is the temporal location of the overload amplitude; $j$ is the index of the identified overloads, and look for the relationship between the correction factor $\lambda_\mathrm{ol}$ and $\nu_\mathrm{ol}$.
Or, equivalently, the PDF of the overload amplitude for all considered cases can be used instead to characterize the rate of occurrence of overload amplitudes, depending on a valid description of the overload amplitude PDF, see \fig{fig:ERol_exp}.
The characterization of the loading rate through the PDF of overload amplitudes is not readily recognizable across all the considered cases.
Therefore, we propose obtaining the correction factor through a functional relation in a data-driven manner, that is, assuming the following functional relation between the correction factor and the moments of the PDF of overload amplitudes,
\begin{equation}
    \lambda_\mathrm{ol} = G(m_\mathrm{ol}^{(1)},m_\mathrm{ol}^{(2)},\dots)
\end{equation}
where $G(\cdot)$ is the functional of overload statistics, $m_\mathrm{ol}^{(i)}, \,i\in\mathbb{Z}^+$ is the $i^\mathrm{th}$ moment of the overload amplitude PDF.
The objective here is to show the feasibility of constructing a simple, functional relationship that captures the irregularity in loading rate that is not described by the fatigue damage estimator in \eq{eq:ettf}.
This procedure should at least improve the prediction accuracy for both experimental and synthetic cases.
A generalized relationship that unifies both observations is desired, if feasible.
To find such a functional relationship in a data-driven manner, a series of methodologies are available from statistical learning~\cite{bishop2006pattern,hastie2009elements}.
We propose using linear regression with \emph{polynomial predictors} derived from the first four moments of the overload amplitude PDF to extract a simple and physically tractable relationship.
With the formulation of the correction factor introduced in detail in the following section, we present, here, the formulation of the proposed fatigue life prediction as,
\begin{equation}
    T^* = \frac{1}{\lambda_\mathrm{ol}\nu_\mathrm{p}C^{-1}\int_{-\infty}^{\infty}p(\sigma_\mathrm{rar})d\sigma_\mathrm{rar}}.
    \label{eq:newT}
\end{equation}
where $\nu_\mathrm{p}$ is the expected rate of occurrence of peaks from the acquired load signals.

\begin{figure}
    \centering
        \includegraphics[width = 0.35\textwidth]{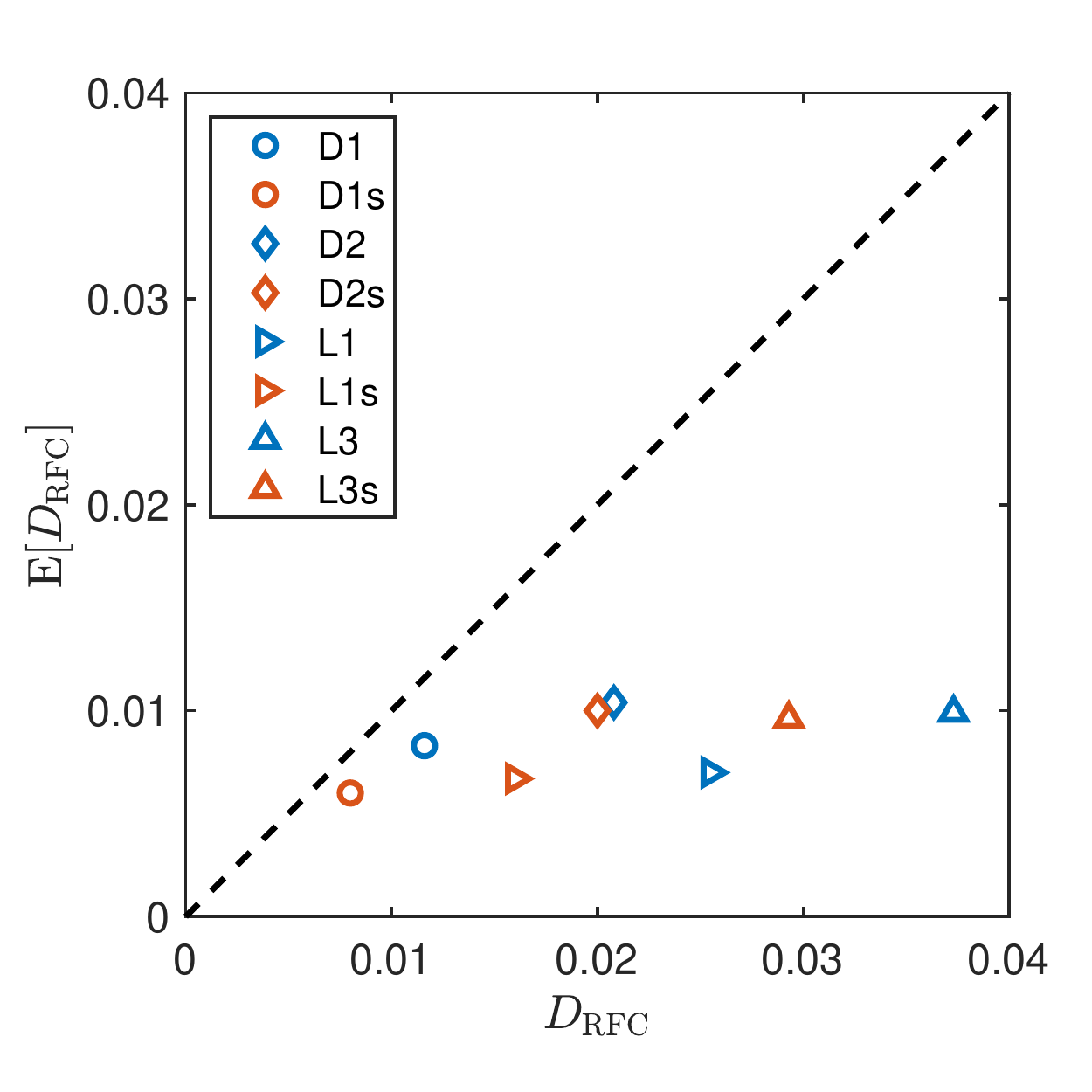}
        \includegraphics[width = 0.35\textwidth]{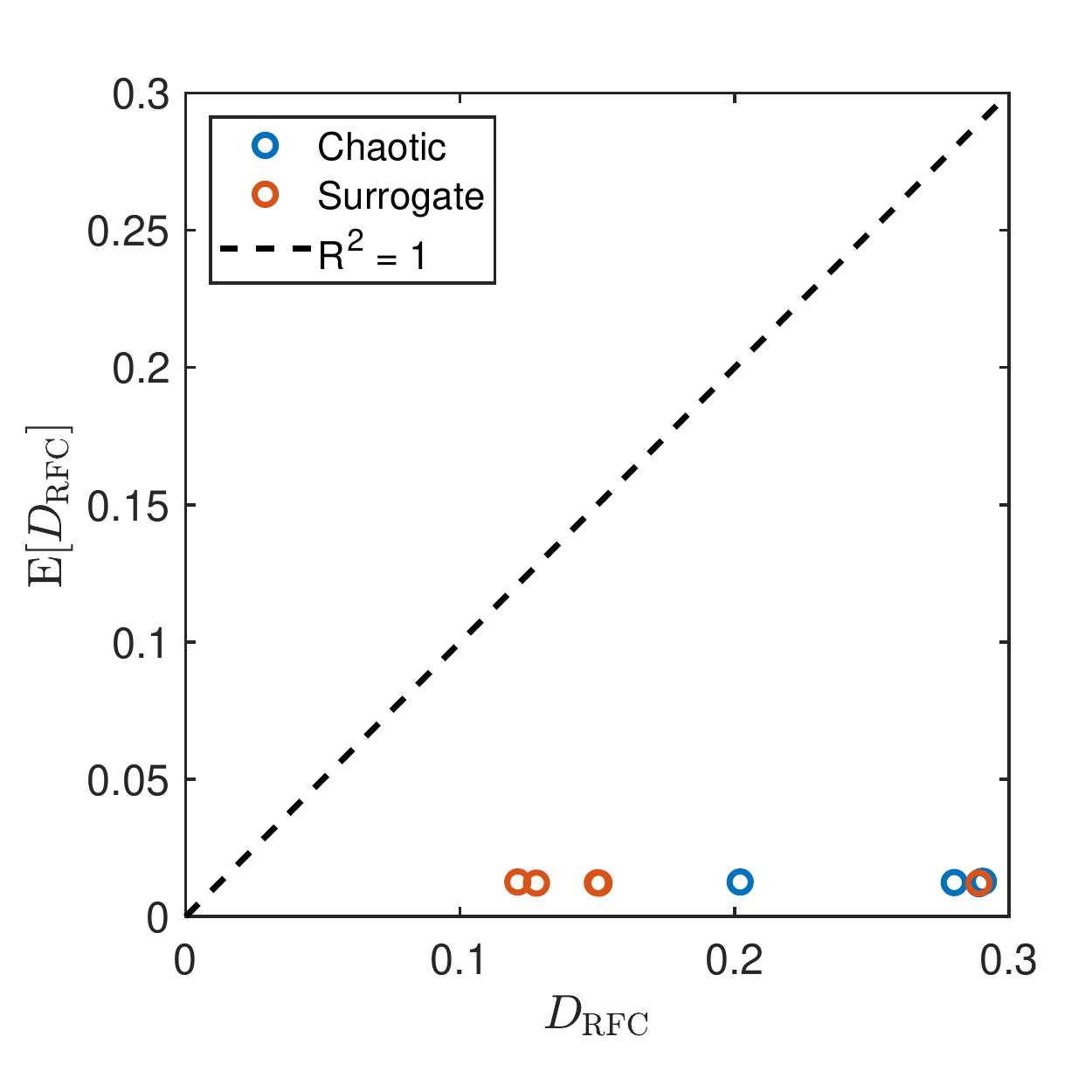}
    \caption{Comparison of fatigue damage estimation between the Miner's rule, (\eq{eq:ettf}, and the spectral method~\cite{benasciutti2005spectral} indicating underestimated damage by the spectral method. Left: generated synthetic load time histories; right: load time histories from the experiments.}
    \label{fig:BT_Drfc}
\end{figure}

\begin{figure}[h]
    \centering
    \includegraphics[width = 0.35\textwidth]{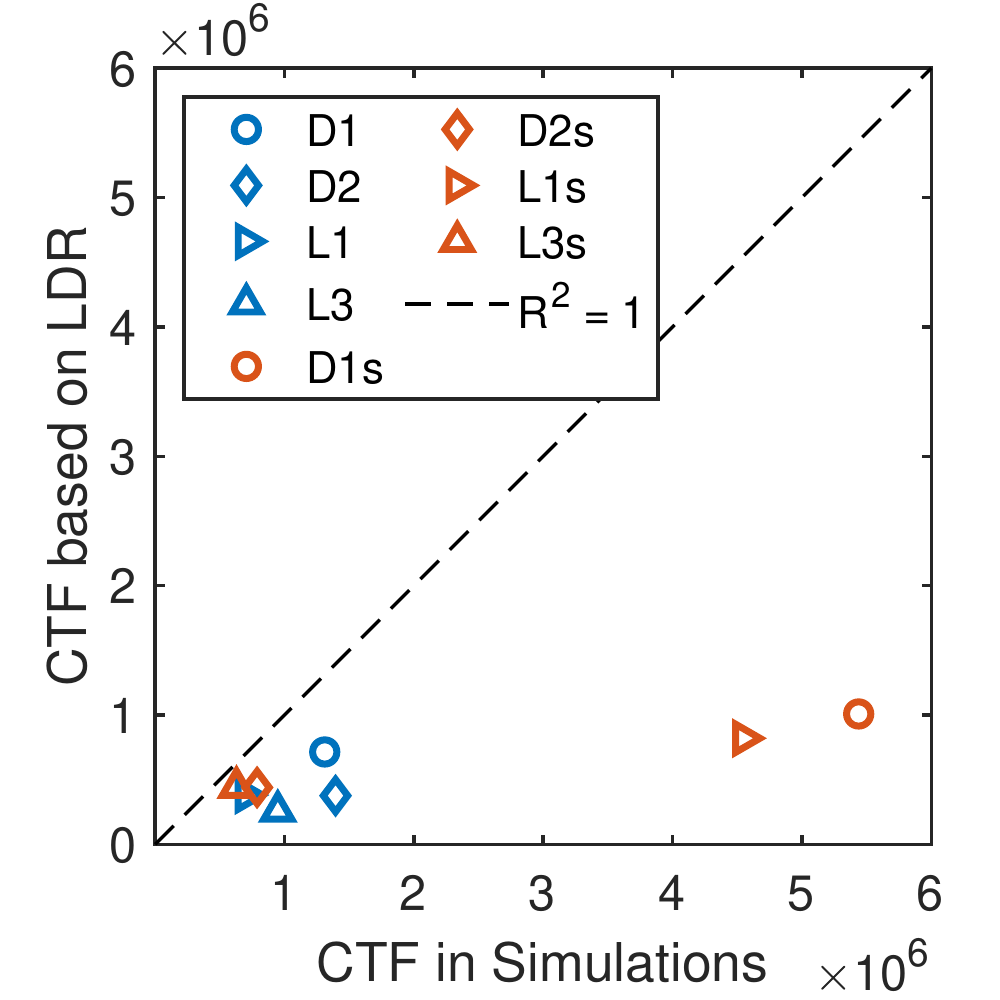}
    \includegraphics[width = 0.35\textwidth]{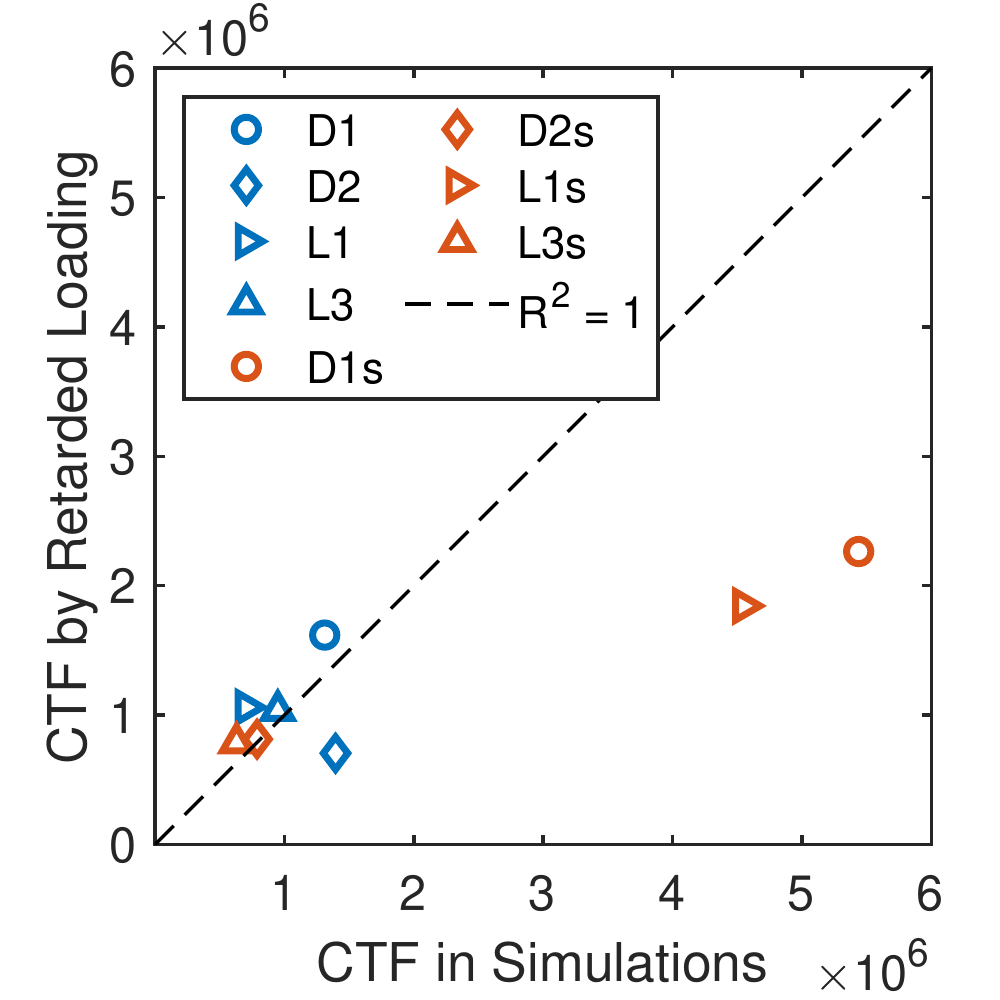}
    \caption{Comparison of fatigue life estimation in terms of CTF between left: Miner's rule with stress amplitude PDF and right: proposed retarded stress amplitude PDF. These results indicate improvement in life prediction accuracy when the overload sequence effect is considered using the proposed model of load sequence effects.}
    \label{fig:CTFwoCorr}
\end{figure}

\begin{figure}[h]
    \centering
    \includegraphics[width = 0.35\textwidth]{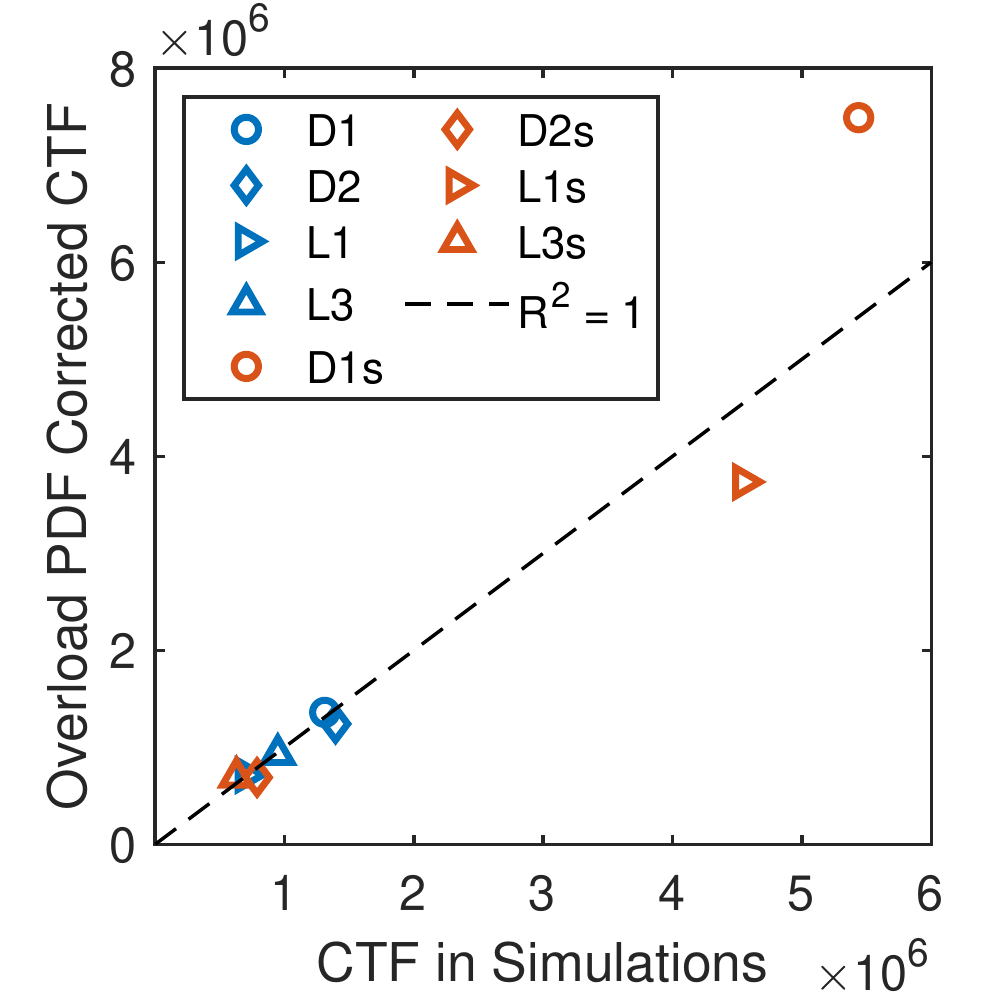}
    \includegraphics[width = 0.35\textwidth]{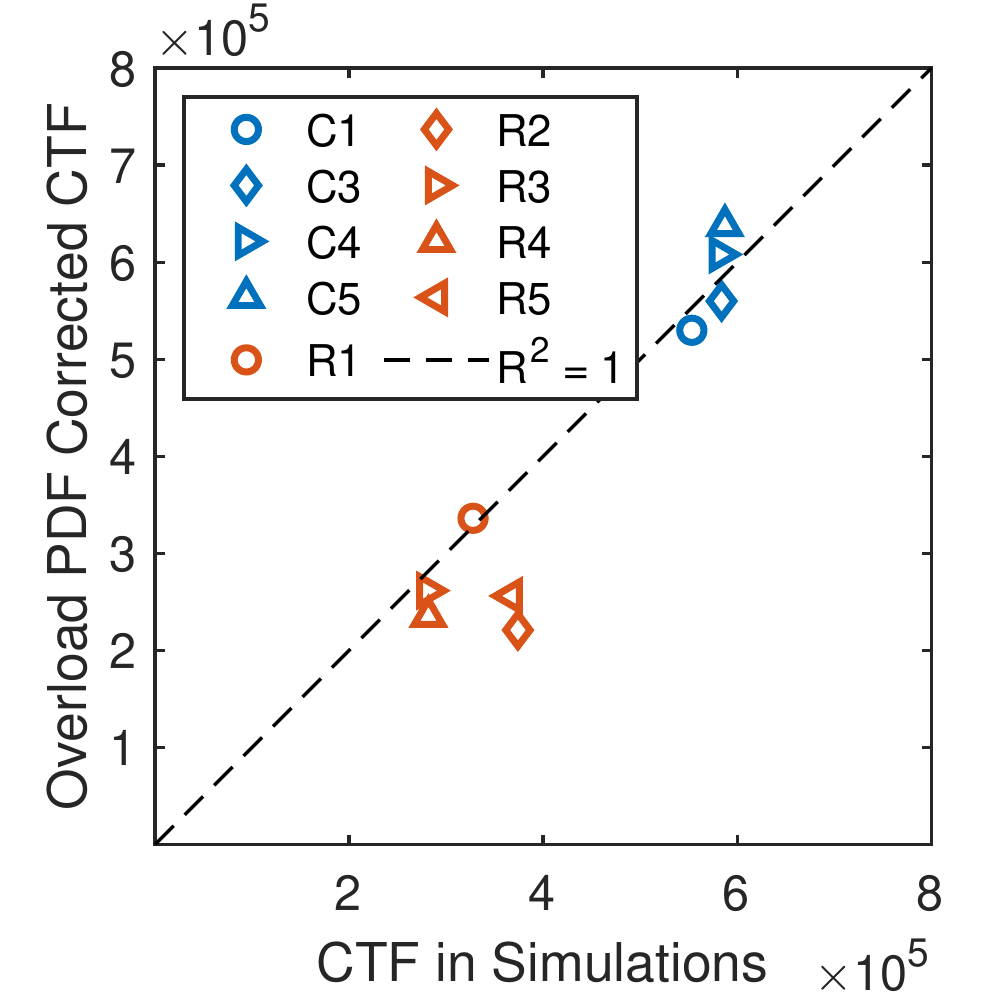}
    \caption{Fatigue life prediction results using the proposed life prediction method with a unified overload correction factor. Left: CTF estimation using synthetic load time histories; right: CTF estimation using experimental load time histories. These results indicate that the data-driven overload rate correction further improves the accuracy of fatigue life estimation under the considered (1) synthetic loading and (2) loading from the physical experiments.}
    \label{fig:CTFwCorr}
\end{figure}
\section{Results and Discussion}\label{Results}
Before presenting fatigue life prediction results using the proposed framework, we provide a brief discussion on damage estimation using the spectral method~\cite{benasciutti2005spectral}. 
This method's prediction result is compared to the damage estimation using Miner's rule and the rainflow-counting stress amplitude distribution as it claims to be equivalent to those for Gaussian loads.
The loads considered here are mean-subtracted (or equivalently, have zero magnitudes at zero frequency) and satisfy the spectral method's requirements.
As illustrated in \fig{fig:BT_Drfc}, the spectral method fails to match the rainflow-counting-based damage estimates for both synthetic and experimental data.
It is no surprise to observe underestimated CTF results since the spectral method, by design, works with Gaussian random processes.
More importantly, the experimentally obtained load time histories are prone to additive noise, which makes the loading a combination of underlying determinism and stochastic processes, which is not necessarily Gaussian.
As a result, the mean up-crossing occurrence rate is not as clearly defined when the random process is non-Gaussian.
Also, since irregularity factors are functions of moments of the PDF of a given random process, the correction cannot delineate the difference between the chaotic loading and their surrogates.  

We examine the life estimation capability of the proposed framework, and present the fatigue life estimation results from both the synthetic and the experimental load time histories and compare them to the Miner's rule.
One can observe from the left plot in \fig{fig:CTFwoCorr}, when fatigue life is estimated using Miner's rule, all synthetic cases are prone to overestimating their damage. 
With the proposed framework without considering the effects from the overloads' statistics, an improved correlation can be observed from the right plot in \fig{fig:CTFwoCorr}, where five of the synthetic cases are being shifted closer to the actual CTF. 
Three other cases seem to overestimate damage by the scale of two (the slope of the line passing through D2, D1s, and D2s cases is half of the other cases).
Therefore, just integrating stress amplitudes does not capture the fatigue damage/fatigue evolution's underlying dynamics.
Thus, a correction factor, which is similar to the irregularity factor, is used to further improve behavior in the estimated life, reflecting the underlying dynamics.
This paper assumes this underlying variable depends on the overload statistics since the damage estimation, calculated from only the integral to the PDF of stress amplitudes, does not include any information from the overload distributions.
In particular, we consider the correction factor a functional form of the first four central moments of the overload amplitude PDF.
The functional relationship between the correction factor and the moments is solved in a least square sense by implementing polynomial regression.
\begin{equation}
    \hat{\boldsymbol{\lambda}}_\mathrm{ol} =  \mathbf{M}\boldsymbol{\beta}, 
\end{equation}
where $\hat{\boldsymbol{\lambda}}_\mathrm{ol}$ is the predicted correction factor for a given set of $n^\mathrm{th}$ polynomial features of moments of overload amplitude PDFs, $m_\mathrm{ol}^{(i)}$, in data matrix $\mathbf{M} = P_n[m_\mathrm{ol}^{(1)} m_\mathrm{ol}^{(2)} m_\mathrm{ol}^{(3)} m_\mathrm{ol}^{(4)}]$ with polynomial mapping $P_n[\cdot]$ and the model vector $\boldsymbol{\beta}$.
For the regression model training, all 160 synthetic fatigue life data is used, and two samples of the experimental data (C1, C3, R1, and R2).
The remaining experimental data sets provide the validation set to show the procedure's prediction power and illustrate this data-driven model's correction capability and generality.
For illustration purposes, the CTFs of synthetic cases are a mean estimate of each 20 cases, while all experimental data cases are shown explicitly.
The corrected results are shown in \fig{fig:CTFwCorr}, which demonstrates the improvements to the overall fatigue life predictions.
Moreover, this correction factor is generalized to experimental results, which indicates a reasonably good generalization without over-fitting the training data set.
A polynomial order of $n = 2$ with only interaction terms, linear terms, and the constant term is chosen for the least-mean-square-error calculated from the validation data set.
The model has more noticeable scatter in a higher CTF range; this may result from sparsity in the observed and simulated CTFs, which can be improved when more cases with a broader range of irregular loading are considered.
The expression for $\boldsymbol{\beta}$ and $\lambda_\mathrm{ol}$ are given in \eq{eq:beta} and \eq{eq:lambdaol}, respectively, in the appendix A3.

\section{Conclusions}
A framework for fatigue life estimation under statistically and spectrally similar variable amplitude loading is introduced as an extension to the cumulative damage rule.
In addition to the PDF of stress amplitudes, it considers the load sequence effect by identifying overloads based on the rainflow-counting algorithm, the characterization of overload retardation effects, and a correction to the stress amplitude PDF based on data-driven modeling.
This framework is applied to both experimentally obtained and synthetically generated load time histories which are chaotic solutions to well-known nonlinear dynamical systems with single-well and double-well oscillation, and their corresponding stochastic surrogates with similar spectral and statistical characteristics.
Its application is not limited to Gaussian random loading, making the damage prognosis possible to more complex loads. 
The results show its capability in predicting fatigue life under statistically and spectrally similar loading where most contemporary CDR-based life estimation methodologies fail to give the desired prediction.
The results also indicate a correlation between the fatigue life under irregular loading to the overload amplitude probability distribution that is not reflected from the typically considered linear spectral statistics and moments of the stress amplitude probability distribution.
Results also indicate that the framework can predict fatigue life with reasonably good precision for load time histories that are experimentally acquired.
It does not require \emph{a priori} knowledge of the applied loading, which makes it a desirable tool for damage prognosis for in-service structures under monitoring, given {\em a priory} the material and structural properties.

For completion, we need to test this framework in a broader range of real-life scenarios, which are not only limited to statistically and spectrally similar loading but also generic variable amplitude loading with a broader range of irregularity.
A future extension of this work to a broader range of irregularity can be pursued by incorporating more loading cases with various temporal and frequency domain characteristics.
Further, a study focusing on the functional relationship between fatigue damage and the corresponding overload amplitude PDF is needed for a more reliable and accurate prognosis.

\section{Acknowledgement}
This study is supported by the National Science Foundation Grants No. 1561960.
\clearpage
\begin{appendix}
\addtocontents{toc}{\protect\setChapterprefix{Appendix }}
\section*{Appendix A.}

\subsection*{A.1\quad Algorithm of iterative surrogate data generation}\label{appendixa1}
\begin{table}[h]
\centering
\begin{tabular}{@{}llllllllllll@{}}
\toprule
\multicolumn{12}{l}{Algorithm of Iterative Surrogate Data Generation}                                                                                                                                                                                                                                                                                                                       \\ \midrule
\textbf{Require:}  & \multicolumn{11}{l}{Given time series, $\{x_n\}$ and convergence criteria}                                                                                            \\
                                    & \multicolumn{11}{l}{Begin with a random shuffle of the chaotic load time history, $\{xp_n^{(0)}\}$.}\\
                                    & \multicolumn{11}{l}{\textbf{while} convergence criteria is not met \textbf{do}}\\
                                    & \multicolumn{11}{l}{\begin{tabular}[c]{@{}l@{}}\quad\textbf{for} $i = 1:N$ \textbf{do}\\ 
                                    \qquad Obtain the phase and amplitude of the FFT of $\{x_n^{(i)}\}$,$\{S_n^{(i)}\}$.\end{tabular}}\\
                                    & \multicolumn{11}{l}{\begin{tabular}[c]{@{}l@{}}\qquad Replace the amplitude of $S_n^{(i)}$ by $\abs{S_n} = \abs{\mathrm{FFT}(x_n)}$. \\ \qquad And obtain the frequency magnitude matched time series $xf_n^{(i)} = \mathrm{IFFT}(S_n^{(i)})$.\end{tabular}}                                                                                                  \\
                                    & \multicolumn{11}{l}{\begin{tabular}[c]{@{}l@{}}\qquad Replace the magnitudes of the time series, $\{xf_n^{(i)}\}$, by the magnitude in the original \\ \qquad chaotic time series. The obtained time series $\{xp_n^{(i)}\}$ whos magnitudes in time are \\ \qquad identical to $\{x_n^{(0)}\}$.\\ 
                                    \quad \textbf{end for}\\ \textbf{end while}\end{tabular}} \\
                                    & \multicolumn{11}{l}{\begin{tabular}[c]{@{}l@{}}\textbf{return} frequency magnitude matched surrogate $\{xf_{n}\}$ and amplitude matched \\ surrogate $\{xp_n\}$.\end{tabular}}                                                                                                                                                           \\ \bottomrule
\end{tabular}
\label{tab:itsurr}
\end{table}

\subsection*{A.2\quad Algorithm of AFGROW crack closure model}\label{appendixa2}

\begin{table}[h]
\centering
\begin{tabular}{@{}llllllllllll@{}}
\toprule
\multicolumn{12}{l}{Algorithm for AFGROW Crack Closure Model Simulation}                                                                                                                           \\ \midrule
  \bfseries{Require:}           & \multicolumn{11}{l}{Given fatigue load spectrum}                          \\
                                &\multicolumn{11}{l}{\textbf{while} crack size $a$ is within designated crack length \textbf{do}}\\
                                & \multicolumn{11}{l}{\quad$R_\mathrm{k}^{(i)} = K_\mathrm{min}^{(i)}/K_\mathrm{max}^{(i)}$ (if $K_\mathrm{min}^{(i)}<K_\mathrm{op}^{(i)}$, $K_\mathrm{min}^{(i)} = K_\mathrm{op}^{(i)})$}\\
                                & \multicolumn{11}{l}{\quad$C_\mathrm{f}^{(i)} = K_\mathrm{min}^{(i)}/K_\mathrm{max}^{(i)}$} \\
                                & \multicolumn{11}{l}{\quad$R^{(i)} = \frac{0.4(C_\mathrm{f0}-1)+\sqrt{F(C_\mathrm{f0},C_\mathrm{f})}}{1.2(1-C_\mathrm{f0})}$}\\
\bfseries{Determine R:}& \multicolumn{11}{l}{
                                \(
                                    \quad R^{(i)} = 
                                    \begin{cases}
                                    C_\mathrm{f}^{(i)}& R^{(i)}>R_\mathrm{h}^{(i)} \,\textrm{or}\, R^{(i)}>C_\mathrm{f}^{(i)}\\
                                    R_\mathrm{K}^{(i)}& R_\mathrm{K}^{(i)} < 0 \,\mathrm{and}\, R_\mathrm{K}^{(i)} \leq R^{(i)}\\
                                    \end{cases}
                                \)}\\
                                & \multicolumn{11}{l}{} \\
\bfseries{Determine} $\boldsymbol{\Delta K:}$& \multicolumn{11}{l}{
                                           \(
                                            \quad \Delta K^{(i)} =    
                                            \begin{cases}
                                                \frac{\Delta K_\mathrm{eff}(1-R)}{1-C_\mathrm{f}} &   R^{(i)}\geq 0  \\   
                                                \frac{\Delta K_\mathrm{eff}}{1-C_\mathrm{f}} &   R^{(i)}<0\\     
                                            \end{cases}
                                      \)}\\ 
                                &\multicolumn{11}{l}{\quad \textbf{if} $\Delta K^{(i)} > \Delta K_\mathrm{Ic} $, \textbf{break}}\\
                                &\multicolumn{11}{l}{\quad \textbf{end if}}\\
                                &\multicolumn{11}{l}{\quad Obtain $\Delta a^{(i)}$ from tabular lookup data base or Paris' Law ($\Delta a^{(i)} = f(\Delta K^{(i)})$).}\\
                                &\multicolumn{11}{l}{\quad $a^{(i)} = a^{(i-1)} + \Delta a^{(i)}$}\\ 
                                &\multicolumn{11}{l}{\quad i = i + 1}\\
                                &\multicolumn{11}{l}{\textbf{end while}}\\ \midrule
\bfseries{Note:}                &\multicolumn{11}{l}{$F(C_\mathrm{f0},C_\mathrm{f})=(0.4(1-C_\mathrm{f0}))^2-4(0.6(1-C_\mathrm{f0})(C_\mathrm{f0}-C_\mathrm{f}))$                     }\\ \bottomrule
\end{tabular}
\end{table}
where $K_\mathrm{min}^{(i)}$ ($K_\mathrm{max}^{(i)}$) is the minimum (maximum) stress intensity factor, $K_\mathrm{op}^{(i)}$ is the crack opening stress intensity factor, and $\Delta K_\mathrm{Ic}$ is the fracture toughness of mode I fracture, $a$ is the crack size, and $C_\mathrm{f}$ is the crack closure factor.
\subsection*{A.3\quad Data-driven model for overload rate correction factor}\label{appendixa3}

The data-driven mode parameters are listed below:
\begin{equation}\label{eq:beta}
    \boldsymbol{\beta} = [-27.4,\,2.83,\,5.82,\,-11.5,\,-1.13,\,-0.468,\,-1.45,\,0.224,\,0.907,\,-0.745,\,9.69].
\end{equation}
And the model can be explicitly expressed as:
\begin{multline}\label{eq:lambdaol}
    \lambda_\mathrm{ol} = -27.4 + 2.83\, m_\mathrm{ol}^{(1)}+5.82\,m_\mathrm{ol}^{(2)}-11.5\,m_\mathrm{ol}^{(3)}-1.13\,m_\mathrm{ol}^{(4)}\\
    -0.468\,m_\mathrm{ol}^{(1)}m_\mathrm{ol}^{(2)}-1.45\,m_\mathrm{ol}^{(1)}m_\mathrm{ol}^{(3)}+0.224\,m_\mathrm{ol}^{(1)}m_\mathrm{ol}^{(4)}+0.907\,m_\mathrm{ol}^{(2)}m_\mathrm{ol}^{(3)}-0.745\,m_\mathrm{ol}^{(2)}m_\mathrm{ol}^{(4)}+9.69\,m_\mathrm{ol}^{(3)}m_\mathrm{ol}^{(4)}.
\end{multline}

\end{appendix}



\clearpage

    \nolinenumbers
\clearpage
\bibliographystyle{IEEEtran}
\bibliography{main}

\end{document}